\documentclass[pra,twocolumn,a4paper,showpacs,superscriptaddress]{revtex4-1}
\usepackage{latexsym}
\usepackage{amsmath}
\usepackage{amsfonts}
\usepackage{graphicx,color}
\usepackage{exscale}
\usepackage{amssymb}
\usepackage{bbold,mathrsfs}
\usepackage{natbib}
\newcommand{\ket}[1]{\ensuremath{|#1\rangle}}
\newcommand{\bra}[1]{\ensuremath{\langle  #1 |}}

\newcommand{\ve}{\varepsilon}

\newcommand{\vro}{\varrho}
\newcommand{\be}{\begin{equation}}
\newcommand{\ee}{\end{equation}}
\newcommand{\ba}{\begin{eqnarray}}
\newcommand{\ea}{\end{eqnarray}}

\newcommand{\mc}[1]{\ensuremath{\mathcal{#1}}}
\newcommand{\bc}{\begin{center}}
\newcommand{\ec}{\end{center}}
\newcommand{\bi}{\begin{itemize}}
\newcommand{\ei}{\end{itemize}}
\newcommand{\mean}[1]{\ensuremath{ \langle #1  \rangle}}

\newcommand{\rf}{\vro_{\text{F}}}
\newcommand{\rft}{\tilde{\vro}_{\text{F}}}

\newcommand{\ain}{A_{\text{in}}}
\newcommand{\bin}{B_{\text{in}}}
\newcommand{\aout}{A_{\text{out}}}
\newcommand{\bout}{B_{\text{out}}}
\newcommand{\aouts}{\tilde{A}_{\text{out}}}
\newcommand{\bouts}{\tilde{B}_{\text{out}}}

\newcommand{\lio}{\mc{L}_{\text{IO}}}

\begin{document}

\title{Dissipative quantum light field engineering}

\author{Martin \surname{Kiffner}}
\affiliation{Clarendon Laboratory, University of Oxford, Parks Road, Oxford OX1 3PU, United Kingdom}

\author{Uwe \surname{Dorner}}
\affiliation{Centre for Quantum Technologies, National University of Singapore, 3 Science Drive 2, Singapore 117543}
\affiliation{Clarendon Laboratory, University of Oxford, Parks Road, Oxford OX1 3PU, United Kingdom}

\author{Dieter \surname{Jaksch}}
\affiliation{Clarendon Laboratory, University of Oxford, Parks Road, Oxford OX1 3PU, United Kingdom}
\affiliation{Centre for Quantum Technologies, National University of Singapore, 3 Science Drive 2, Singapore 117543}


%
\begin{abstract} 
We put  forward a dissipative preparation scheme for  strongly correlated photon states. 
Our approach is based on a two-photon loss mechanism that is realised via a 
single four-level atom inside a bimodal optical cavity. 
Each elementary two-photon emission event  removes one photon out of 
each of the two modes. The dark states of this loss mechanism are given by NOON 
states and arbitrary superpositions thereof. We find that the steady state 
of the two cavity modes exhibits entanglement  and for certain 
parameters, a mixture of two coherent entangled states is produced. 
We discuss how the quantum correlations in the cavity modes and the output fields can be measured.  
\end{abstract}

\maketitle

\section{INTRODUCTION} 
The  performance of optical technologies such as metrology, communication and imaging 
can be improved beyond the limitations of classical physics if non-classical light sources are employed.  
The major challenge for the realisation of these quantum-enhanced schemes is 
the deterministic generation of custom-tailored photon states for specific applications.  
For example, quantum information schemes based on continuous variable 
entanglement~\cite{braunstein:05} have the advantage that entangled 
light fields can be generated unconditionally, but high-quality 
resources with a large degree of entanglement are difficult to produce. 
Another example is given by optical interferometry, a technique that  
is employed in  applications like gravitational wave detectors, 
laser gyroscopes or optical imaging. 
It usually aims at the precise estimation of a relative phase acquired by the 
light on its way through the interferometer. 
The achievable precision of the phase estimation with classical light sources 
is bound by the standard quantum limit (SQL) and  
scales as $1/\sqrt{N}$, where $N$ is the (mean) number of photons in the input field~\cite{dorner:09}.  
On the contrary, a better precision with the same amount of resources 
can be obtained with entangled light fields.   
A Prominent example for photon states that allow one to beat 
the SQL is given by  so-called NOON states~\cite{boto:00,bollinger:96}, which also give 
rise to phase super-resolution~\cite{dowling:08}. 
Realistic scenarios that include photon losses 
of the interferometer require states with a more complicated  structure than 
NOON states  for breaking the SQL~\cite{dorner:09,kacprowicz:10,escher:11,thomaspeter:11}.  
Another example is given by coherent entangled states~\cite{sanders:92,sanders:11} (CES)  
that allow one to beat the SQL in lossless interferometers~\cite{gerry:01,gerry:10}. 
These states yield a better phase estimation than NOON states 
in lossy interferometers~\cite{joo:11} if the same mean number of input photons are 
taken into account. 
However,  it remains challenging to produce NOON~\cite{mitchell:04,afek:10} and 
CES~\cite{ourjoumtsev:06,ourjoumtsev:07,joo:11} states  
with a high success rate and with a large number of photons.

One of the
most successful techniques for the preparation of quantum mechanical systems 
in a desired state is dissipation. Prominent examples are given by laser-
and evaporative cooling that allow one to realize a Bose-Einstein 
condensate. 
Recently, the concept of dissipative quantum state preparation~\cite{kraus:08,verstraete:09,diehl:08} 
was transferred  to the many-body domain 
where dissipation alone prepares  strongly correlated states. 
The challenge in dissipative quantum state preparation is 
to design a suitable dissipative process $\mc{L}_{\Gamma}$ such 
that the desired  state $\ket{\psi}$ is  stationary with respect to  $\mc{L}_{\Gamma}$, 
i.e.,  $\mc{L}_{\Gamma}(\ket{\psi}\bra{\psi})=0$. 
For example,  a dissipative contact interaction was 
investigated in one-dimensional molecular~~\cite{syassen:08,ripoll:08,duerr:09,daley:09}
and polariton~\cite{kiffner:10,kiffner:10b,kiffner:11} systems, where  
dissipation effectively results in a repulsion between particles. 
The  entanglement of two distant atomic ensembles via spontaneous emission 
was investigated in~\cite{krauter:10b,muschik:10b}, and the simulation of 
open quantum systems with ion systems was considered in~\cite{barreiro:11,mueller:11}. 

Here we present a dissipative preparation scheme for  strongly correlated photon states 
inside an optical cavity with two modes $a$ and $b$. We engineer a 
two-photon loss term  via  a single,  
laser-driven four-level atom that couples to the cavity modes, see Fig.~\ref{picture1}. 
Each elementary emission event induced  by this two-photon loss term removes one photon out of 
mode $a$ and one photon out of mode $b$.  
The dark states of the engineered dissipator $\mc{L}_{\Gamma}$ 
are given by  all strongly entangled NOON states and superpositions thereof, 
and the stationary state inside the cavity 
can be well approximated by a mixture of two CES states for specific parameters.   
We show that the steady state of the cavity modes alone is entangled, and 
point out how the entanglement of the cavity modes and of the output field 
can be measured. 
This paper is organised as follows. We give a detailed description of 
our system (see Fig.~\ref{picture1})   in Sec.~\ref{model}. Here we also 
derive an effective master equation for the cavity modes alone, which we 
obtain by an  adiabatic elimination of the atomic degrees of freedom. 
In Sec.~\ref{steady}  we analyse the steady state of this effective  master equation 
and identify the dark states of the engineered two-photon loss mechanism. 
The entanglement of the two cavity modes is discussed in Sec.~\ref{entanglement}. 
We employ the Negativity~\cite{vidal:02} and an inequality~\cite{duan:00}  
based on Einstein-Podolsky-Rosen type observables as sufficient entanglement criteria. 
The predictions of both measures are compared, and Sec.~\ref{output} indicates how 
the criterion based on the inequality~\cite{duan:00} could be measured experimentally. 
The latter Sec.~\ref{output} is mostly concerned with the entanglement of the output field. 
We identify two suitable modes of the output field whose entanglement can 
be inferred from two-mode squeezing spectra. 
Finally, we address the experimental realisation of our scheme in Sec.~\ref{realisation} 
and conclude with  a summary and outlook of our results in Sec.~\ref{conclusion}. 
\section{REDUCED MASTER EQUATION FOR THE CAVITY MODES \label{model}}
\begin{figure}[t!]
\includegraphics[scale=1]{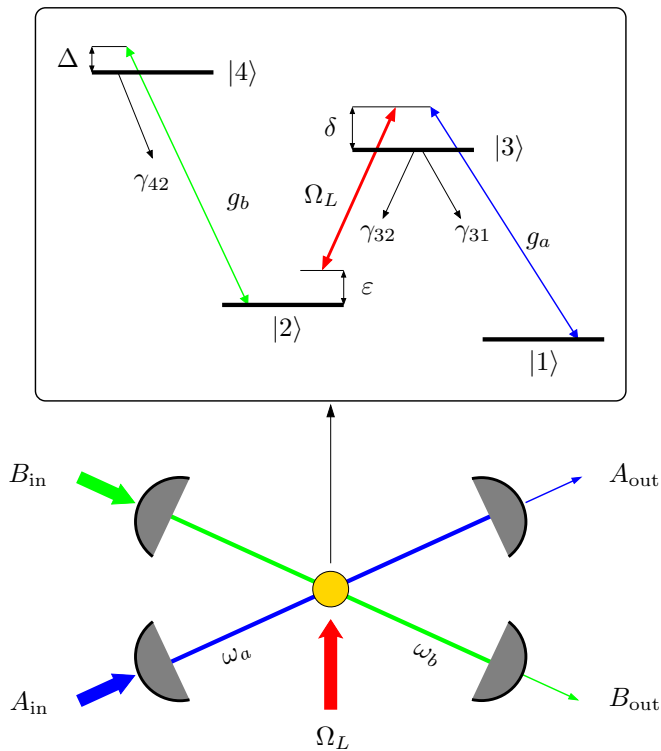}
\caption{\label{picture1} \small   
(Color online) A single four-level atom 
interacts with two cavity modes and a classical laser field.  
$\ain$ and $\bin$ are coherent input fields that drive the 
corresponding cavity resonantly, and $\aout$, $\bout$ are the 
output fields.  
The inset shows the atomic level scheme.  
The  laser field with frequency $\omega_{\text{L}}$ and 
Rabi frequency $\Omega_{\text{L}}$ couples to the $\ket{2}\leftrightarrow\ket{3}$ 
transition, and the cavity mode with frequency $\omega_a$  interacts with the $\ket{3}\leftrightarrow\ket{1}$ 
transition. The second cavity mode with frequency $\omega_b$ interacts with the $\ket{4}\leftrightarrow\ket{2}$ 
transition. $g_a$ and $g_b$ are the single-photon Rabi frequencies corresponding to mode  $a$ and $b$, respectively.  
The parameters $\gamma_{ij}$ are the  decay rates of the various transitions, $\delta$ and $\Delta$ 
label the detuning of the cavity fields with transition $\ket{3}\leftrightarrow\ket{1}$ and 
$\ket{4}\leftrightarrow\ket{2}$, respectively, and $\ve$ is the two-photon detuning.  
 }
\end{figure}
We consider a single four-level atom that  
interacts with two  cavity modes and a classical 
laser  field, see Fig.~\ref{picture1}.  
It was shown theoretically~\cite{schmidt:96} and experimentally~\cite{kang:03} that 
the considered level scheme can give rise to a strongly enhanced Kerr nonlinearity. 
In addition, this level configuration allows one to engineer a two-photon absorption 
process~\cite{harris:98,kiffner:10b}. 
Note that we chose the two-cavity setup in Fig.~\ref{picture1} because it allows 
us to present our model in a clear and unambiguous way. 
However, our setup could very well be realised with a single cavity where 
the two  modes can be either two different polarisation or frequency modes.  
In particular, the two modes could have the same frequency and orthogonal polarisations. 

The aim of this section is to derive 
an equation of motion for the reduced density operator $\rf$ of the 
two cavity modes. 
We begin with a detailed description of the system shown in 
Fig.~\ref{picture1}. The input field $\ain$ ($\bin$) is a coherent field 
that resonantly drives the cavity mode with frequency $\omega_a$ ($\omega_b$). 
The corresponding Hamiltonian is 
\be
H_{\text{in}} =  \hbar\Omega_a^*  e^{i\omega_a t} a + \hbar \Omega_b^*  e^{i\omega_b t} b + \text{H.c.},
\ee
where H.c. stands for the Hermitian conjugate, and 
$a$ $(b)$  is the annihilation  operator of 
the cavity mode with frequency $\omega_a$ ($\omega_b$). 
The Rabi frequencies $\Omega_a$ and $\Omega_b$ are determined by 
the input power of the fields $\ain$ and $\bin$, respectively. 
The cavity mode with frequency $\omega_a$ couples to the atomic transition 
$\ket{3}\leftrightarrow\ket{1}$, and the  mode with frequency $\omega_b$ 
interacts with the atom on the $\ket{4}\leftrightarrow\ket{2}$ transition. 
In rotating-wave approximation (RWA), the interaction of the atom 
with the cavity modes  is described by the Hamiltonian 
\be
H_{\text{C}} =  -\hbar  g_a a \ket{3}\bra{1} -  \hbar  g_b b \ket{4}\bra{2} + \text{H.c.} ,
\ee 
where $g_a$ ($g_b$) is the single-photon Rabi frequency on the $\ket{3}\leftrightarrow\ket{1}$
($\ket{3}\leftrightarrow\ket{2}$) transition. 
The detuning of the first cavity mode with the $\ket{3}\leftrightarrow\ket{1}$ 
transition is denoted by $\delta$, and $\Delta$ is the detuning of the 
second mode with the  $\ket{4}\leftrightarrow\ket{2}$ transition, 
\be
\delta = \omega_a - \omega_{31}\, , \qquad \Delta=\omega_b - \omega_{42} \,. 
\label{detuning}
\ee
The resonance frequencies on the $\ket{3}\leftrightarrow\ket{1}$ and 
$\ket{4}\leftrightarrow\ket{2}$ transitions have been labelled by $\omega_{31}$ 
and $\omega_{42}$, respectively. 
In addition, the atom interacts with a classical laser field with frequency $\omega_{\text{L}}$ and 
Rabi frequency $\Omega_{\text{L}}$, and this field couples to the $\ket{3}\leftrightarrow\ket{2}$ 
transition. In rotating-wave approximation, the atom-laser interaction reads 
\be
H_{\text{L}} =  -\hbar \Omega_{\text{L}} \ket{3}\bra{2} e^{-i\omega_{\text{L}} t} + \text{H.c.} . 
\ee 
The  free time evolution of the cavity modes and  of the atomic degrees of freedom is 
given by $H_{\text{F}}$ and $H_{\text{A}}$, respectively, 
\begin{align}
& H_{\text{F}} =  \hbar\omega_a a^{\dagger}a + \hbar \omega_b b^{\dagger}b , \label{hf} \\
& H_{\text{A}} =  \hbar[ \omega_2 \ket{2}\bra{2} + \omega_3 \ket{3}\bra{3} + \omega_4 \ket{4}\bra{4}], \label{ha}
\end{align}
and we  set $\omega_1=0$ in Eq.~(\ref{ha}). 
With these definitions, we arrive at the master equation 
for the combined system of the atomic degrees 
of freedom  and the two cavity modes, 
\be
\dot{\vro} = - \frac{i}{\hbar} [ H_{\text{F}}+H_{\text{in}}  + H_{\text{A}}  + H_{\text{L}} + H_{\text{C}}, \vro ] 
+\mc{L}_{\gamma}\vro + \mc{L}_{\kappa}\vro \,.
\label{master_eq}
\ee
The  term $\mc{L}_{\gamma}\vro$ in Eq.~(\ref{master_eq}) describes spontaneous emission 
of  the atom and is given by 
\begin{align}
\mc{L}_{\gamma}\vro = & -\frac{\gamma_{31}}{2}
\left( S_1^+S_1^- \vro + \vro S_1^+S_1^-  - 2 S_1^- \vro S_1^+ \right) \notag \\
& -\frac{\gamma_{32}}{2}
\left( S_2^+S_2^- \vro + \vro S_2^+S_2^-  - 2 S_2^- \vro S_2^+ \right) \notag \\
& -\frac{\gamma_{42}}{2}
\left(S_3^+S_3^- \vro + \vro S_3^+S_3^-  - 2 S_3^- \vro S_3^+  \right) ,\label{decay0}
\end{align}
where $\gamma_{ij}$  is the full decay rate on the transition $\ket{i}\leftrightarrow\ket{j}$ 
(see Fig.~\ref{picture1}). 
The atomic transition operators are defined as 
\be
S_1^+ =\ket{3}\bra{1}  ,\quad S_2^+ = \ket{3}\bra{2} ,\quad S_3^+ =\ket{4}\bra{2} ,
\ee
and $S_i^- = (S_i^+)^{\dagger}$.
The last term $\mc{L}_{\kappa}\vro$ in Eq.~(\ref{master_eq}) accounts for photon losses 
at the cavity mirrors and reads
\begin{align}
\mc{L}_{\kappa}\vro = & -\frac{\kappa_a}{2}(a^{\dagger}a\vro + \vro a^{\dagger}a -2 a \vro a^{\dagger}) \notag \\
& -\frac{\kappa_b}{2}(b^{\dagger}b\vro +\vro b^{\dagger}b -2 b \vro b^{\dagger}), \label{lk}
\end{align}
where $\kappa_a$ ($\kappa_b$) is the damping rate of mode $a$ ($b$). 

In appendix~\ref{derivation}, we derive from Eq.~(\ref{master_eq}) 
the master equation for the reduced density operator $\rf$ of the cavity modes alone, 
\be
\rf= \text{Tr}_{\text{A}}\vro = 
\vro_{11}+\vro_{22}+\vro_{33}+\vro_{44}\,,
\ee
and $\vro_{\nu\nu}$ denotes $\bra{\nu}\vro\ket{\nu}$. 
%
%
If the two-photon detuning $\ve= \omega_a - \omega_{\text{L}} -\omega_2$ 
vanishes, the master equation for the density operator $\rft$ of the cavity modes in 
an interaction picture with respect to $H_{\text{F}}$ in Eq.~(\ref{hf}) is given by 
\be
\dot{\tilde{\vro}}_{\text{F}}  = \mc{L}_{\text{in}}\rft +\mc{L}_{\kappa}\rft 
+\mc{L}_{U}\rft+\mc{L}_{\Gamma}\rft, 
\label{master_field}
\ee
where 
\be
\mc{L}_{\text{in}}\rft = - i[\Omega_a^* a+\Omega_b^* b+\Omega_a a^{\dagger}+\Omega_b b^{\dagger},\rft ]
\label{lin}
\ee
accounts for the coherent driving of the cavity modes and the cavity decay term 
$\mc{L}_{\kappa}\rft$ is defined in Eq.~(\ref{lk}). 
The two terms $\mc{L}_{U}\rft$ and $\mc{L}_{\Gamma}\rft$ in Eq.~(\ref{master_field}) 
represent the influence of the single atom on the evolution of the two cavity modes. 
More specifically, 
\be
\mc{L}_{U}\rft = - i \, U [(b a)^{\dagger}b  a, \rft] 
\label{Lu}
\ee
describes a coherent two-particle interaction between photons in modes $a$ and $b$. 
The strength of this interaction is determined via the parameter
\be
U = \frac{|g_a|^2 |g_b|^2}{(\Delta^2+\gamma_{42}^2/4)|\Omega_{\text{L}}|^2}\Delta.
\ee
Note that the sign of $U$ depends on the sign of the detuning $\Delta$ defined in Eq.~(\ref{detuning}). 
It follows that the interaction described by $\mc{L}_{U}$ can be repulsive or attractive. 
The last term in Eq.~(\ref{master_field}) is given by 
\be
\mc{L}_{\Gamma}\rft =  -\frac{\Gamma}{2}\left[(ba)^{\dagger}ba\rft 
+ \rft (ba)^{\dagger}ba -2 ba \rft (ba)^{\dagger}\right] \label{GammaD}
\ee
and represents a two-photon loss term with decay rate  
\be
\Gamma = \frac{|g_a|^2 |g_b|^2}{(\Delta^2+\gamma_{42}^2/4)|\Omega_{\text{L}}|^2} \gamma_{42}. 
\label{Gamma}
\ee
The dissipator  $\mc{L}_{\Gamma}$ gives rise to the emission of correlated photon pairs.
In each elementary emission process,  $\mc{L}_{\Gamma}$ removes one photon 
out of mode $a$ and one photon out of mode $b$. 
In the following, we focus on  the quantum correlations between the cavity modes 
that are induced by the two-photon loss term  $\mc{L}_{\Gamma}$. 
For the rest of this paper, we will thus consider the purely 
dissipative scenario where the conservative photon-photon 
interaction vanishes, i.e. $U=0$.  This situation arises 
when mode $b$ is resonant with the $\ket{4}\leftrightarrow\ket{2}$ transition and thus $\Delta=0$. 

Next we summarise the conditions for the validity of the reduced master equation~(\ref{master_field}). 
The adiabatic elimination of the atomic degrees of freedom requires that the atomic 
decay rates $\gamma_{ij}$ are large as compared to the other system parameters,
\be
\gamma_{42},\gamma_{31},\gamma_{32} \gg \kappa_a,\kappa_b, \sqrt{N_a N_b} \Gamma,  
\Omega_a, \Omega_b,|\delta|, 
\label{adcon}
\ee
where $N_a$ ($N_b$) is the largest relevant photon number in mode $a$ ($b$). 
The master equation~(\ref{master_field}) for the cavity modes holds under the assumption 
that the two-photon detuning $\ve$ vanishes. 
Since $\ve=0$ can be adjusted only within a certain accuracy, we establish 
conditions that ensure the validity of Eq.~(\ref{master_field}) for $\ve\not=0$ in Appendix~\ref{sec_cond}. 
Finally, the restriction to terms up to third order in the expansion of the 
density operator~(\ref{expansion}) requires 
\be
\frac{N_a |g_a|^2 }{|\Omega_{\text{L}}|^2 } \ll 1,
 \frac{N_b |g_b|^2 }{|\Omega_{\text{L}}|^2 } \ll 1,
\label{cond_x}
\ee
where $N_a$ ($N_b$) is the maximal photon number in mode $a$ ($b$). 
\section{STEADY STATE ANALYSIS \label{steady}}
%
\begin{figure*}[t!]
\begin{center}
\includegraphics[scale=1]{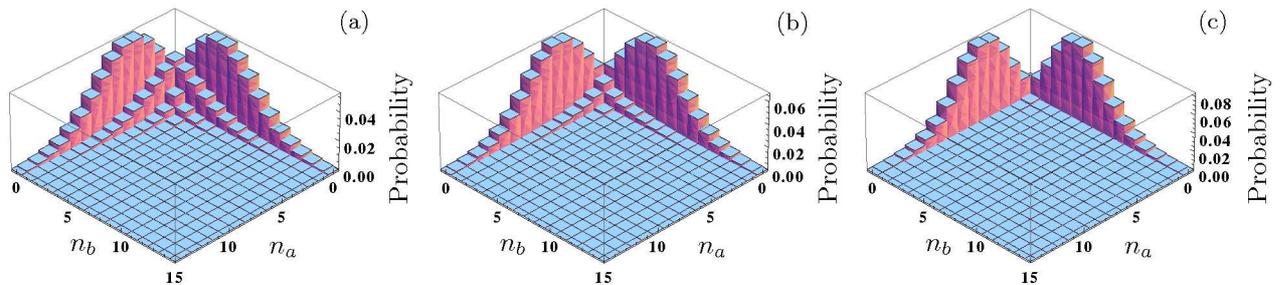}
\caption{\label{picture5} \small (Color online) Population of the cavity Fock states $\ket{n_a,n_b}$ in 
steady state for 
(a) $\Omega_a=\Omega_b=1.32\times 10^{-2}\gamma_{42}$, $\kappa_a=\kappa_b=10^{-2}\gamma_{42}$, 
(b) $\Omega_a=\Omega_b=6.26\times 10^{-3}\gamma_{42}$, $\kappa_a=\kappa_b=0.5\times 10^{-2}\gamma_{42}$,  and 
(c) $\Omega_a=\Omega_b=1.16\times 10^{-3}\gamma_{42}$, $\kappa_a=\kappa_b=10^{-3}\gamma_{42}$.
The common parameters in (a)-(c) are $\Gamma=\gamma_{42}/100$, $\Delta=U=0$, $\delta=0$ and $\ve=0$. 
The mean photon number for all states in (a)-(c) is $\mean{\hat{N}}=5$. 
}
\end{center}
\end{figure*}
%
%
In this Section we characterise the steady state of the master equation~(\ref{master_field}) 
with a purely dissipative two-photon interaction ($\Delta=U=0$). 
Throughout this work we assume that the 
 cavity decay rates $\kappa_a$ and $\kappa_b$ are different from zero. In this case 
and for time-independent input fields $\Omega_a$ and $\Omega_b$,  
the master equation~(\ref{master_field}) exhibits a unique steady state. 
If the two-photon decay term $\Gamma$ vanishes, this unique steady state is the pure state~\cite{kraus:08} 
$\rft=\ket{\psi}\bra{\psi}$, where 
$\ket{\psi}=\ket{\alpha_a,\alpha_b}$ and $\ket{\alpha_a}$ ($\ket{\alpha_b}$) denotes a coherent state in 
mode $a$ ($b$) with amplitude $\alpha_{a} =-2i\Omega_a/\kappa_a$ ($\alpha_b=-2i\Omega_b/\kappa_b$).

Next we turn to the general situation with $\Gamma\not=0$. In this case, 
the analytical steady state of the master equation~(\ref{master_field})  is difficult to obtain.  
A numerical study   shows that the steady state of Eq.~(\ref{master_field}) is in general a 
mixed state.   In addition, an intuitive understanding of the structure of this state can be 
gained via the dark states~\cite{kraus:08} $\ket{D}$ of the two-photon loss term 
in Eq.~(\ref{GammaD}) that obey $\mc{L}_{\Gamma}(\ket{D}\bra{D})=0$. 
Since $\mc{L}_{\Gamma}$  removes one photon out of each cavity mode in every 
elementary emission event, all pure dark states $\ket{D}$ 
must be of the form
\be
\ket{D} = c_1 \ket{\phi_a,0} + c_2 \ket{0,\varphi_b}, 
\label{darkstate}
\ee
where $\ket{\phi_a}$ and $\ket{\varphi_b}$ are arbitrary states (including the vacuum) 
of mode $a$ and $b$, respectively. 
Note that $\mc{L}_{\Gamma}$ supports an infinite number of dark states, and 
the most general dark state is  given by a mixture of different pure dark states 
that are of the form of $\ket{D}$ in Eq.~(\ref{darkstate}).
In the regime where  the two-photon decay rate $\Gamma$ dominates over 
the coherent drive terms $\Omega_a$, $\Omega_b$ and the cavity decay rates $\kappa_1$, $\kappa_2$, 
one can expect that the stationary state of Eq.~(\ref{master_field}) is approximately 
a dark state of $\mc{L}_{\Gamma}$ alone.  
This is confirmed by Fig.~\ref{picture5} that shows the population of the cavity Fock states $\ket{n_a,n_b}$ in 
steady state  for various parameters. These results were obtained via a numerical solution of Eq.~(\ref{master_field}).  
The relative importance of the two-photon loss term~(\ref{GammaD}) is the smallest 
in Fig.~\ref{picture5}(a) and the largest in Fig.~\ref{picture5}(c). In the latter case 
it is apparent that the steady state is comprised of the dark states in Eq.~(\ref{darkstate}). 
Only Fock states $\ket{0,n_b}$ and $\ket{n_a,0}$ are significantly populated, while 
the population of other states $\ket{n_a,n_b}$ with $n_a\not=0$ and $n_b\not=0$ is strongly 
suppressed.
This result can be understood as follows. All Fock states 
$\ket{n_a,n_b}$ with $n_a\not=0$ and $n_b\not=0$ experience not only the cavity loss term $\mc{L}_{\kappa}$, 
but the additional (strong) two-particle losses $\mc{L}_{\Gamma}$. This suppresses the population of 
these states via the cavity input fields. 

The cavity pump fields  $\Omega_a$, $\Omega_b$ induce transitions between 
dark states $\ket{n_a,0}$, $\ket{0,n_b}$ and neighbouring Fock states $\ket{n_a,1}$, $\ket{1,n_b}$ 
 ($n_a\not=0$ and $n_b\not=0$).  The latter states are not dark states 
of $\mc{L}_{\Gamma}$ and thus decay rapidly. 
However, we point out that this mechanism does not induce an indirect decay 
of the dark states $\ket{0,n_b}$, $\ket{n_a,0}$ if $\Gamma$ becomes much larger than all other system parameters. 
The reason is that  the transitions $\ket{0,n_b}\rightarrow\ket{1,n_b}\rightarrow\ket{0,n_b-1}$ 
($\ket{n_a,0}\rightarrow\ket{n_a,1}\rightarrow\ket{n_a-1,0}$) occur at an effective  
rate~\cite{comment1}  $\Omega_a^2/\Gamma\ll 1$ ($\Omega_b^2/\Gamma\ll 1$)
for $\Omega_a\ll \Gamma$ ($\Omega_b\ll \Gamma$) and are therefore negligible. 
This result can be regarded as a manifestation of  the quantum Zeno effect~\cite{ripoll:08} 
and explains the sharp population contrast between  dark states $\ket{n_a,0}$, $\ket{0,n_b}$ 
and neighbouring Fock states $\ket{n_a,1}$, $\ket{1,n_b}$ in Fig.~\ref{picture5}(c).

The most general pure dark state in Eq.~(\ref{darkstate}) is a coherent superposition 
of the states $\ket{\phi_a,0}$ and $\ket{0,\varphi_b}$. This is a remarkable 
feature since it implies that  all strongly entangled NOON states 
\be
\ket{\text{NOON}}=\frac{1}{\sqrt{2}}(\ket{N,0}+\ket{0,N})
\ee
and coherent superpositions thereof are dark states of the dissipator $\mc{L}_{\Gamma}$. 
Next we investigate whether the  steady state of Eq.~(\ref{master_field}) contains 
entangled dark states or just a mixture of the states $\ket{\phi_a,0}$ and $\ket{0,\varphi_b}$. 
Note that this information is not contained in Fig.~\ref{picture5} because it does not contain 
any information about the off-diagonal matrix elements of the density operator. 
We find that the steady state  exhibits non-vanishing 
coherences in the Fock basis if the modes $a$ and $b$ appear in a completely 
symmetric fashion in the  master equation~(\ref{master_field}). 
For example, for the parameters in Fig.~\ref{picture5}(c), the diagonalisation of 
the numerically computed  density operator shows  that the steady state is very well 
approximated by  a mixture of two coherent entangled states (CES), 
\begin{align}
\rft\approx & p_{1}\ket{\text{CES}_+(\alpha_1)}\bra{\text{CES}_+(\alpha_1)} \notag \\
& + p_{2}\ket{\text{CES}_-(\alpha_2)}\bra{\text{CES}_-(\alpha_2)} ,
\label{approximate}
\end{align}
where $p_{1}\approx0.499$, $p_{2}\approx0.459$, $\alpha_1\approx- 2.23 i$, $\alpha_2\approx- 2.29 i$ and 
the CES states are defined as~\cite{sanders:92,joo:11} 
\begin{align}
\ket{\text{CES}_{\pm}(\alpha)}=\frac{1}{\sqrt{2(1 \pm e^{-|\alpha|^2})}}(\ket{\alpha,0}\pm\ket{0,\alpha}). 
\end{align}
The overlap between the approximate state in Eq.~(\ref{approximate}) and the full 
numerical solution in the trace norm is  $95.44\%$. 
Recently, the performance of CES in interferometric precision measurements was 
discussed in~\cite{joo:11}. Since it was found that these states perform  better than 
NOON states in a lossy interferometer, we discuss the suitability of the state 
in Eq.~(\ref{approximate}) for interferometric precision measurements.  
We find that the mixture in Eq.~(\ref{approximate}) performs approximately as good 
as classical light with a well-defined phase. Note, however, that the 
performance of a pure CES and the mixture in Eq.~(\ref{approximate}) differ only 
if  the losses of the interferometer are smaller than approximately 25\%. In this 
regime of a near-perfect interferometer, pure CES entangled states allow one 
to access the quantum regime that is currently inaccessible with our preparation method. 
The previous discussion indicates that the steady state of the  two cavity 
modes can be prepared in an entangled state. The entanglement properties of the cavity modes 
and of the output field are  discussed in the following Sections~\ref{entanglement} 
and~\ref{output}, respectively. 
\section{ENTANGLEMENT OF THE CAVITY FIELD \label{entanglement}}
The two modes of the cavity form a bipartite quantum system. 
By definition, the quantum state $\rf$ of the cavity field is said to 
be entangled if and only if  it is non-separable, and 
$\rf$  is separable if and only if it can be written as
\be
\rf = \sum\limits_j p_j \vro_a(j) \otimes \vro_b(j)\,. 
\ee
Here  $\vro_a(j)$ and $\vro_b(j)$ are  normalised states 
of the  modes $a$ and $b$, respectively, and the parameters  
$p_j\ge 0$ are constrained by $\sum_j p_j =1$.  

We employ two criteria that are both sufficient for 
the entanglement of the cavity modes. We begin with 
the  Negativity~\cite{vidal:02} that is defined as 
\be
\mc{N}(\rf) = \frac{1}{2}\left(1-\lVert\rf^{\text{T}_a}\rVert_1\right),
\label{neg}
\ee
where $\rf^{\text{T}_a}$ denotes the partial transpose of $\rf$ 
with respect to the subsystem $a$. The trace norm 
$\lVert A \rVert_1 = \text{Tr}(\sqrt{A^{\dagger}A})$ of an operator $A$ 
is equal to the sum of its singular values. Note that the 
Negativity is an entanglement monotone, and hence 
$\mc{N}(\rf)>0$ is a sufficient criterion for the entanglement of the two 
field modes. 

%
\begin{figure}[t!]
\includegraphics[width=8cm]{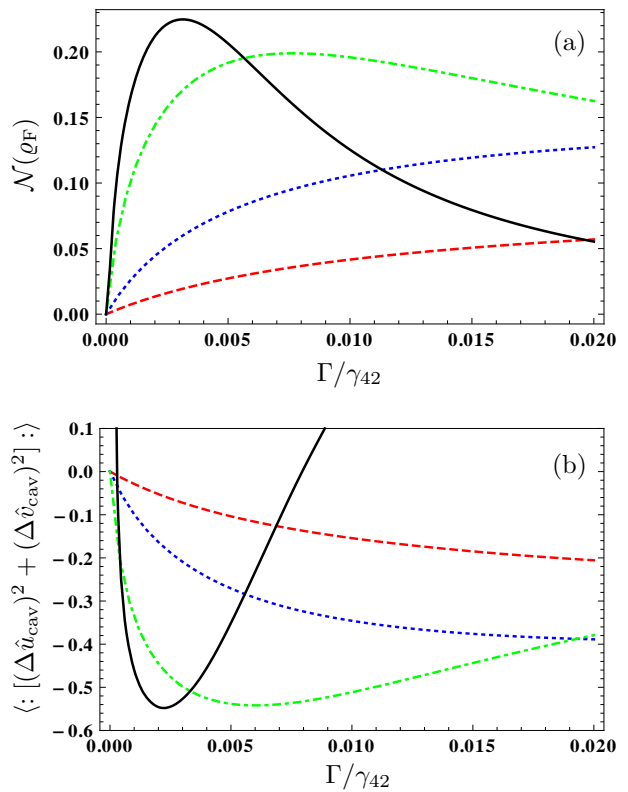}
\caption{\label{picture2} \small (Color online) 
Numerical evaluation of (a) the  Negativity and (b) the inequality 
in Eq.~(\ref{criterion2}) for the two cavity modes 
in steady state. 
The results are shown as a function of the 
two-particle loss rate $\Gamma$ and for different intensities of the 
coherent driving fields.  
The parameters are $\phi=\pi/2$, $\kappa_a=\kappa_b=10^{-2} \gamma_{42}$, 
$\Delta=U=0$, $\delta=0$ and $\ve=0$. The red dashed line corresponds to 
$\Omega_a=\Omega_b=0.2\times 10^{-2}\gamma_{42}$, the blue dotted line stands for 
$\Omega_a=\Omega_b=0.4\times 10^{-2}\gamma_{42}$, the green dashed-dotted line corresponds to  
$\Omega_a=\Omega_b=1\times 10^{-2}\gamma_{42}$ and the black solid line stand for 
$\Omega_a=\Omega_b=1.6\times 10^{-2}\gamma_{42}$. 
}
\end{figure}
%
The second criterion is derived in~\cite{duan:00} and 
states that the system is in an entangled quantum state if   
the total variance of two Einstein-Podolsky-Rosen (EPR) type   
operators $\hat{u}$ and $\hat{v}$ of the two modes satisfy the inequality   
\be
\mean{\left(\delta\hat{u}_{\text{cav}}\right)^2  + \left(\delta\hat{v}_{\text{cav}}\right)^2} < 2 \,,
\label{criterion}
\ee
where 
\be
\hat{u}_{\text{cav}} = \hat{x}_a + \hat{x}_b \,,\qquad \hat{v}_{\text{cav}} = \hat{p}_a - \hat{p}_b \,.
\label{u_and_v}
\ee
For any operator $X$, we define $\delta \hat{X} = \hat{X} - \mean{\hat{X}}$, and 
$\hat{x}_k$ and $\hat{p}_k$ are local operators which 
correspond to mode $k\in\{a,b\}$ with frequency $\omega_k$. 
The only restriction imposed on the operators $\hat{x}_k$ and $\hat{p}_k$ is that 
they  must obey the commutation relation
\be
[\hat{x}_k, \hat{p}_l] = i\delta_{kl}\,, 
\label{com}
\ee
but are otherwise arbitrary. Here we choose for $\hat{x}_k$ and $\hat{p}_l$ 
the quadrature operators of the cavity fields 
\begin{align}
&\hat{x}_a = \frac{1}{\sqrt{2}}(\tilde{a} e^{-i \phi} + \tilde{a}^{\dagger}e^{i \phi}) ,  \label{quadraturesA}\\
&\hat{p}_a = \frac{1}{\sqrt{2} i} (\tilde{a} e^{-i \phi} - \tilde{a}_k^{\dagger}e^{i \phi}), \\
&\hat{x}_b = \frac{1}{\sqrt{2}}(\tilde{b} e^{-i \phi} + \tilde{b}^{\dagger}e^{i \phi}) ,  \\
&\hat{p}_b = \frac{1}{\sqrt{2} i} (\tilde{b} e^{-i \phi} - \tilde{b}_k^{\dagger}e^{i \phi}), 
\label{quadraturesB}
\end{align}
where $\tilde{a}= e^{i\omega_a t} a$ and $\tilde{b}= e^{i\omega_b t} b$ are 
slowly varying in time. 
Since $\hat{u}_{\text{cav}}$ and $\hat{v}_{\text{cav}}$ can be identified with the 
operators corresponding to the centre-of-mass motion and the 
relative momentum of two quantum mechanical oscillators, respectively, they are called EPR type operators. 
The phase $\phi$ is arbitrary and will be optimised such that a maximal violation of the 
inequality~(\ref{criterion}) is achieved. 

The inequality in Eq.~(\ref{criterion}) is equivalent to 
\be
\mean{:[\left(\delta\hat{u}_{\text{cav}}\right)^2  + \left(\delta\hat{v}_{\text{cav}}\right)^2]:} < 0 \,,
\label{criterion2}
\ee
where $:\ :$ denotes normal ordering (all creation operators to the left) 
with respect to the operators $a$, $b$ and their adjoints. For reasons that will become clear 
later (see Sec.~\ref{output}), we will employ Eq.~(\ref{criterion2}) instead of Eq.~(\ref{criterion}). 
With the help of Eqs.~(\ref{u_and_v}) and~(\ref{quadraturesA})-(\ref{quadraturesB}), we find
\begin{align}
& \mean{: [(\delta\hat{u}_{\text{cav}})^2  + (\delta\hat{v}_{\text{cav}})^2 ]:}  = 
 2\big[
\mean{\delta \tilde{a}^{\dagger}\delta \tilde{a}} \notag\\
&\hspace*{1cm}  + \mean{\delta \tilde{b}^{\dagger}\delta \tilde{b}} + \mean{\delta \tilde{a}\delta \tilde{b}} e^{-2i\phi} 
+ \mean{\delta\tilde{a}^{\dagger} \delta\tilde{b}^{\dagger}}  e^{2i\phi} \big]. 
\label{lhs_criterion}
\end{align}

We numerically solve for the steady state of the density operator of the two cavity modes via 
the master equation~(\ref{master_field}) for various parameters $\Gamma$ and different 
intensities of the coherent input fields.  The result for the  entanglement criteria in Eqs.~(\ref{neg}) 
and~(\ref{lhs_criterion}) is shown in Figs.~\ref{picture2}(a) and~(b), respectively. 
It follows that the steady state of the system  exhibits entanglement. 
The Negativity and the normally ordered variance of the EPR-type operators show 
a qualitatively similar (but mirrored) behaviour in the weak driving regime. 
A different situation arises if the Rabi frequencies 
$\Omega_a$ and $\Omega_b$ become comparable to the cavity decay rates. 
Although the cavity modes are  entangled for all parameters  in Fig.~\ref{picture2}(a) 
(the Negativity is larger than zero), 
the sufficient entanglement criterion in Eq.~(\ref{criterion2}) is not fulfilled for 
larger values of $\Gamma/\gamma_{42}$. 
We show in Sec.~\ref{output} and Appendix~\ref{squeeze} that the 
quantity in Eq.~(\ref{lhs_criterion}) and hence 
the entanglement criterion in Eq.~(\ref{criterion2}) can be measured experimentally. 
The discussion above shows that this approach 
allows one to capture the entanglement for a large range of parameters, but it 
fails to detect the entanglement in some cases.
\section{ENTANGLEMENT OF THE OUTPUT FIELD \label{output}}
The output field of the cavity is a multi-mode field. The question whether the output 
field is entangled thus requires to specify the corresponding modes. 
In standard input-output theory~\cite{gardiner:85,clerk:10}, 
the output fields $\aout$ and $\bout$ [see Fig.~\ref{picture1}] are defined as
\begin{align}
& \aout(t)=\frac{1}{\sqrt{2 \pi}} \int \text{d}\omega e^{-i \omega (t-t_1)} A(\omega,t_1),  \\
& \bout(t)=\frac{1}{\sqrt{2 \pi}} \int \text{d}\omega e^{-i \omega (t-t_1)} B(\omega,t_1).
\end{align}
Here $A(\omega,t_1)$ and $B(\omega,t_1)$ are the Heisenberg operators of the continuous output modes 
taken at time $t_1\rightarrow\infty$. The latter operators  obey the equal-time commutation relations
\begin{align}
 & [A(\omega,t),A^{\dagger}(\omega^{\prime},t)]=[B(\omega,t),B^{\dagger}(\omega^{\prime},t)]
=\delta(\omega-\omega^{\prime}). 
\label{commutation}
\end{align}
Here we focus on  the entanglement between two modes corresponding to the central frequencies 
$\omega_a$ and $\omega_b$ of $\aout$ and $\bout$, respectively. Our aim is to follow a similar 
approach as in  Eq.~(\ref{criterion2}) where we employ the variance of EPR-type operators as 
a sufficient criterion for entanglement~\cite{duan:00}. Therefore, we have to define position- and momentum-like 
operators that obey the canonical commutation relation~(\ref{com}). In order to achieve this, 
we have to construct a discrete mode of $\aout$ from the continuous mode operator $A(\omega,t_1)$. 
This transition  can be achieved if we average $A(\omega,t_1)$ 
over a small frequency interval  $\Delta\omega$ centered at $\omega_a$, 
\begin{align}
 \mc{A}_0 = & \frac{1}{\sqrt{\Delta\omega}}
\int\limits_{-\Delta\omega/2}^{\Delta\omega/2} \text{d}\omega 
e^{i (\omega_a +\omega) t_1} A(\omega_a +\omega,t_1),\\
= & \frac{1}{\sqrt{2 \pi \Delta\omega}}
\int\limits_{-\Delta\omega/2}^{\Delta\omega/2} \text{d}\omega
\int\limits_{-\infty}^{\infty}\text{d}t \aouts(t) e^{i\omega t}, \notag
\end{align}
where the output field $\aouts(t)=e^{i \omega_a t}\aout(t)$  is slowly varying in time. 
Similarly, we define a discrete mode of the output field $\bout$ centered at $\omega_b$,  
\begin{align}
\mc{B}_0 
= & \frac{1}{\sqrt{2 \pi \Delta\omega}}
\int\limits_{-\Delta\omega/2}^{\Delta\omega/2} \text{d}\omega
\int\limits_{-\infty}^{\infty}\text{d}t \bouts(t) e^{i\omega t}, 
\end{align}
and $\bouts(t)=e^{i \omega_b t}\bout(t)$. 
The modes $\mc{A}_0$ and $\mc{B}_0$ represent modes of the output fields 
$\aout$ and $\bout$ that are experimentally accessible via spectral filtering. 
Furthermore, we note that $\mean{\mc{A}_0^{\dagger}\mc{A}_0}$ 
can be interpreted~\cite{agarwal:qst2} as the mean number of 
photons emitted into the mode $\mc{A}_0$ within the time $1/\Delta\omega$ 
(the same statement holds for the mode $\mc{B}_0$). 

Since the two discrete modes  $\mc{A}_0$ and 
$\mc{B}_0$ obey the commutation relations
\begin{align}
& [\mc{A}_0,\mc{A}_0^{\dagger}] = [\mc{B}_0,\mc{B}_0^{\dagger}]=1, 
\quad [\mc{A}_0,\mc{B}_0^{\dagger}] = [\mc{A}_0,\mc{B}_0]=0, 
\end{align}
we can define position- and momentum-like operators for each mode 
that obey the canonical  commutation relation~(\ref{com}),  
\begin{align}
&\hat{x}_{\mc{A}} = \frac{1}{\sqrt{2}}(\mc{A}_0 e^{-i \phi} + \mc{A}_0^{\dagger}e^{i \phi}) , 
\label{quadratures2A}  \\
&\hat{p}_{\mc{A}} = \frac{1}{\sqrt{2} i} (\mc{A}_0 e^{-i \phi} - \tilde{A}_0^{\dagger}e^{i \phi}) \\
&\hat{x}_{\mc{B}} = \frac{1}{\sqrt{2}}(\mc{B}_0 e^{-i \phi} + \mc{B}_0^{\dagger}e^{i \phi}) ,  \\
&\hat{p}_{\mc{B}} = \frac{1}{\sqrt{2} i} (\mc{B}_0 e^{-i \phi} - \mc{B}_0^{\dagger}e^{i \phi}). 
\label{quadratures2D}
\end{align}
The phase $\phi$ in Eqs.~(\ref{quadratures2A}) and~(\ref{quadratures2D}) 
is arbitrary and later on chosen such that the necessary separability 
condition is  maximally violated. 
We can now proceed as in Sec.~\ref{entanglement} and define EPR-type operators 
\be
\hat{u}_{0} = \hat{x}_{\mc{A}} + \hat{x}_{\mc{B}} \,,\qquad 
\hat{v}_{0} = \hat{p}_{\mc{A}} - \hat{p}_{\mc{B}}\,.
\label{u_and_v_2}
\ee
With the help of Eqs.~(\ref{u_and_v_2}) and~(\ref{quadratures2A})-(\ref{quadratures2D}),   
we find that 
the normally ordered total variance of the operators $\hat{u}_{0}$ and $\hat{v}_{0}$ is given by
\begin{align}
 \mean{: [(\delta\hat{u}_{0})^2  + (\delta\hat{v}_{0})^2 ]:}    =  
\frac{1}{\Delta\omega}\int\limits_{-\Delta\omega/2}^{\Delta\omega/2}
\text{d}\omega [S_u(\omega) + S_v(\omega)],
\label{crit3}
\end{align}
where $S_u(\omega)$ and $S_v(\omega)$ are related to the two-mode squeezing spectra 
of the output fields. A definition of  these functions and how they can be 
calculated numerically  and measured experimentally is provided in Appendix~\ref{squeeze}. 
%
\begin{figure}[t!]
\includegraphics[width=8cm]{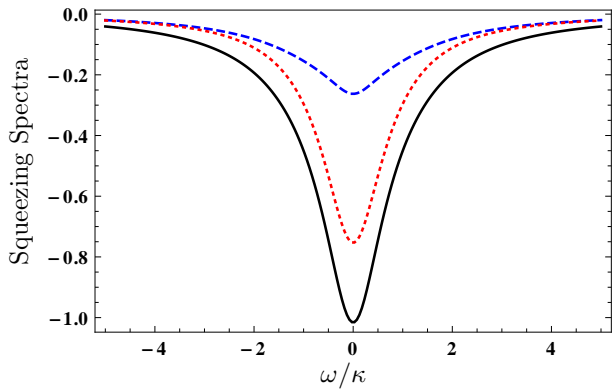}
\caption{\label{picture3} \small (Color online) 
Squeezing spectra. The blue dashed line corresponds to $S_u(\omega)$, the 
red  dotted line to $S_v(\omega)$ and the black solid line shows $S_u(\omega)+S_v(\omega)$. 
The parameters are $\phi=\pi/2$, $\kappa=\kappa_a=\kappa_b=10^{-2} \gamma_{42}$, 
$\Gamma = \gamma_{42}/100$, $\Omega_a=\Omega_b=0.7\times 10^{-2}\gamma_{42}$, 
$\Delta=U=0$, $\delta=0$ and $\ve=0$. 
}
\end{figure}
%
According to the entanglement criterion in~\cite{duan:00}, the modes $\mc{A}_0$ 
and $\mc{B}_0$ are entangled if the left-hand side of Eq.~(\ref{crit3}) is 
smaller than zero. If $\Delta\omega$ is sufficiently small, this means 
that modes $\mc{A}_0$ and $\mc{B}_0$ are entangled if $S_u(0) + S_v(0)< 0$. 
The numerical evaluation of the squeezing spectra $S_u$ and $S_v$ in Fig.~\ref{picture3} 
demonstrates clearly the entanglement of the output modes. For the chosen 
parameters, we find a two-mode squeezing of $-1.3\text{dB}$ in $S_u$ and 
$-6.1\text{dB}$ in $S_v$ at the central frequency $\omega=0$.

So far we considered the entanglement between the two central frequency components 
at $\omega_a$ and $\omega_b$. Note that our approach can be generalised 
to modes centered around $\omega_a-\omega$ and $\omega_b+\omega$ if the output 
spectra of modes $a$ and $b$ are completely symmetric and if the two modes are 
interchangeable in the master equation~(\ref{master_field}). 
In previous works~\cite{braunstein:05,morigi:06,morigi2:06}, other EPR-type 
operators than those in Eqs.~(\ref{u_and_v_2}) 
with~(\ref{quadratures2A})-(\ref{quadratures2D}) were defined, 
and  their variances can also be related to squeezing spectra. 
However, in the latter approach it is not obvious to identify the 
modes of the output field that are entangled. 

Finally, we address the experimental verification of the entanglement inside the cavity. 
To this end, we note that 
\be
\frac{1}{\kappa}\int\limits_{-\infty}^{\infty}
\text{d}\omega [S_u(\omega) + S_v(\omega)] = 
 \mean{: [(\delta\hat{u}_{\text{cav}})^2  + (\delta\hat{v}_{\text{cav}})^2 ]:},
\ee
and hence it follows that the field inside the cavity is entangled if the sum of 
$S_u(\omega) + S_v(\omega)$, integrated over all frequencies, is negative [see Eq.~(\ref{criterion2})]. 
Note that the variance of the cavity fields can be measured directly without 
recording the squeezing spectra~\cite{jing:06}. 
\section{EXPERIMENTAL REALIZATION  \label{realisation}}
Next we discuss the  experimental implementation of our scheme in Fig.~\ref{picture1}.  
A key requirement  of our scheme is that  the single atom should have a very well defined position 
that changes very little over the range of an optical wavelength. 
Ideal candidates are therefore single trapped neutral atoms~\cite{weber:09,khudaverdyan:09} 
or ions~\cite{stute:11a,russo:09,mundt:03,keller:04,devoe:96} inside an optical cavity. 
Recently, several experiments with ${}^{40}\text{Ca}^{+}$ ions inside a 
high-finesse optical cavity have been reported~\cite{stute:11a,russo:09,mundt:03,keller:04}. 
A suitable candidate would also be  given by a ${}^{138}\text{Ba}^{+}$ ion~\cite{devoe:96}, 
where the level scheme in Fig.~\ref{picture1} could be realised between two $J=1/2$ Zeeman 
manifolds  via polarisation- and frequency selection. 
Alternatively, the level scheme in Fig.~\ref{picture1} can be realised with 
artificial atoms~\cite{rebic:09} coupled to microwave fields.

In our  approach we obtain an effective, engineered master equation for the 
cavity modes alone via an adiabatic elimination of the atomic degrees of freedom. 
In particular, this requires that that the cavity decay rates are much smaller than 
the atomic spontaneous emission decay rates, see Eq.~(\ref{adcon}). 
In a recent experiment~\cite{stute:11a,russo:09} with a ${}^{40}\text{Ca}^{+}$ 
ion in a high-finesse cavity, the ratio between the atomic decay rate and the 
corresponding cavity decay rate is $\gamma/\kappa\approx 16$. 
At the same time, the coupling constant $g$  was larger than the atomic decay rate, 
$g/\gamma \approx 2$~\cite{stute:11a}.  However, we show in appendix~\ref{para} that 
it is advantageous to realise our scheme with    $g/\gamma < 1$. 
This opens up the possibility to increase the length $L$  of the cavity 
which will reduce the magnitudes of the coupling constant  $|g|\sim 1/\sqrt{L}$ 
and of the cavity decay rate $\kappa\sim 1/L$. From the above example, we conclude 
that values of  $\gamma/\kappa\approx 100$ should be achievable with 
current technology. 
Valid choices for the  remaining system parameters 
that comply with the conditions~(\ref{adcon}) and~(\ref{cond_x}) of our model 
 are discussed in appendix~\ref{para}. 
In summary, an increase in the number of photons in the cavity  requires 
a reduction in  the two-photon decay rate $\Gamma$. 
Note that this does not necessarily result in a smaller entanglement between the 
cavity modes. On the contrary, Fig.~\ref{picture2} indicates that the maximal entanglement 
is attained for smaller two-photon decay rates $\Gamma$ if the number of photons increases 
(see Sec.~\ref{entanglement}). 
\section{SUMMARY AND OUTLOOK \label{conclusion}}
This paper investigates a scheme for dissipative quantum state preparation for 
two optical cavity modes $a$ and $b$. 
We engineer an effective reservoir via a single, laser-driven four level atom 
that interacts with both cavity modes on separate transitions. 
The adiabatic elimination of the atomic degrees of freedom gives rise 
to a master equation for the cavity modes alone and contains a 
two-photon loss term.  
Each elementary emission event induced  by this two-photon loss term removes one photon out of 
mode $a$ and one photon out of mode $b$. We find that the dark states of 
this loss term are given by  all entangled NOON states and superpositions thereof.   
If the two cavity modes are interchangeable in the corresponding master equation, the 
steady state of the cavity field exhibits entanglement. We employ the Negativity as well as 
an inequality~\cite{duan:00}  
based on Einstein-Podolsky-Rosen type observables as sufficient entanglement criteria. 
While the former is an entanglement monotone, the latter can be measured via two 
balanced homodyne detection setups of the output fields. 
Furthermore, we define suitable modes of the output fields and show that they can 
be entangled as well. The entanglement of the output modes can be verified 
experimentally via the measurement of two-mode squeezing spectra.

The stationary state inside the cavity 
can be well approximated by a mixture of two CES states for specific parameters. 
While a pure CES state is able to access the quantum regime in interferometric 
precision measurements with small photon losses inside the interferometer, 
the mixture prepared by our dissipative scheme does not perform better than 
classical light with a well defined phase. 
However, modifications of our scheme may pave the way towards the 
dissipative preparation of  NOON states and CES that are relevant for 
quantum-enhanced technologies. 
In the ideal case, it may help to prepare photon states that 
allow one to achieve the ultimate quantum limit~\cite{escher:11} in 
interferometric precision measurements with  realistic photon losses.

\begin{acknowledgments}
MK was supported by a fellowship within the Postdoc-Programme of the 
German Academic Exchange Service (DAAD).
\end{acknowledgments}
\appendix
\section{Derivation of the reduced master equation \label{derivation}}
Here we outline the derivation of the effective master equation for the cavity modes in Eq.~(\ref{master_field}). 
The starting point for our derivation is the full master equation in Eq.~(\ref{master_eq}). 
Next we apply a unitary  transformation $W=W_{\text{F}} \otimes W_{\text{A}}$ to 
Eq.~(\ref{master_eq}), where $W_{\text{F}} =  \exp[i H_{\text{F}} t/\hbar]$  acts only on 
the cavity modes, and 
\begin{align}
W_{\text{A}} = & \exp[i( H_{\text{A}}/\hbar + \ve \ket{2}\bra{2} + \delta \ket{3}\bra{3}
+ (\Delta+\ve)  \ket{4}\bra{4}) t]
\end{align}
acts only on the atomic degrees of freedom. 
The density operator in the new frame is 
denoted by $\tilde{\vro} = W\vro W^{\dagger}$ and 
obeys the equation of motion
\be
\dot{\tilde{\vro}} = - \frac{i}{\hbar} [ H_0  + H_{\text{C}},\tilde{\vro} ]
 +\mc{L}_{\gamma}\tilde{\vro}+ \lio\tilde{\vro}\,,
\label{master_eq2}
\ee
where 
\begin{align}
H_0 =  -\hbar[ & \ve \ket{2}\bra{2} + \delta  \ket{3}\bra{3} + (\Delta +\ve) \ket{4}\bra{4} \notag \\
& + \Omega_{\text{L}} \ket{3}\bra{2} + \Omega_{\text{L}}^* \ket{2}\bra{3}] ,
\end{align}
and $\ve= \omega_a - \omega_{\text{L}} -\omega_2$ is the two-photon detuning. 
The super-operator $\lio = \mc{L}_{\text{in}} + \mc{L}_{\kappa}$ 
in Eq.~(\ref{master_eq2}) accounts for the external driving via the input fields  
and the damping of the cavity modes, where $\mc{L}_{\text{in}}$ 
and $\mc{L}_{\kappa}$ are defined in Eqs.~(\ref{lin}) and~(\ref{lk}), respectively. 
The master equation for the transformed density operator 
$\tilde{\vro}_{\text{F}}$ of the cavity modes  is obtained if 
we trace over the atomic degrees of freedom in Eq.~(\ref{master_eq2}), 
\be
\dot{\tilde{\vro}}_{\text{F}}  = \left( i  g_a^*  [ a^{\dagger}, \tilde{\vro}_{31} ] 
+ i g_b^*  [ b^{\dagger}, \tilde{\vro}_{42}  ]   +\text{H.c.}\right) + \lio\rft .
\label{master_field_W}
\ee
In order to eliminate the  coherences $\tilde{\vro}_{31}$ and $\tilde{\vro}_{42}$ 
from Eq.~(\ref{master_field_W}), we  solve Eq.~(\ref{master_eq2}) 
perturbatively  in the Hamiltonian $H_{\text{C}}$ describing the interaction 
between the atom and the cavity modes.  
In order to obtain the desired expansion of the full density operator 
$\tilde{\vro}$ in the coupling constants $g_a$ and $g_b$,  
we re-write Eq.~(\ref{master_eq2}) in 
a form where the atom-cavity interaction is separated from the other terms, 
\be 
\dot{\tilde{\vro}} =\mc{L}_0 \tilde{\vro}  - \frac{i}{\hbar} [ H_{\text{C}},\tilde{\vro} ] ,
\label{master_eq3}
\ee
and the super-operator $\mc{L}_0$ is defined by 
\be
\mc{L}_0 \tilde{\vro} = -  \frac{i}{\hbar} [ H_0 , \tilde{\vro} ] +\mc{L}_{\gamma}\tilde{\vro} 
+ \lio\tilde{\vro}.
\ee
Expansion of the density operator in Eq.~(\ref{master_eq3}) as 
\be
\tilde{\vro} = \sum\limits_{k=0}^{\infty} \tilde{\vro}^{(k)}, 
\label{expansion}
\ee
where $\tilde{\vro}^{(k)}$ denotes the contribution to $\tilde{\vro}$ in 
$k$th order in $H_{\text{C}}$, leads to the following set of 
coupled differential equations 
\begin{align}
&\dot{\tilde{\vro}}^{(0)} =  \mc{L}_0 \tilde{\vro}^{(0)} \,, \label{rho_0} \\
&\dot{\tilde{\vro}}^{(k)} =  \mc{L}_0 \tilde{\vro}^{(k)} -\frac{i}{\hbar}[H_{\text{C}} ,\tilde{\vro}^{(k-1)}] ,
\quad k>0.
\label{iterative}
\end{align}
Equation~(\ref{rho_0}) describes  the interaction of 
the atom with the classical laser fields to all orders and 
in the absence of the cavity fields. 
Higher-order contributions to $\tilde{\vro}$ can be obtained 
if Eq.~(\ref{iterative}) is solved iteratively. 
Equations~(\ref{rho_0}) and~(\ref{iterative}) must be solved under the constraints
$\text{Tr}(\tilde{\vro}^{(0)})=1$ and $\text{Tr}(\tilde{\vro}^{(k)})=0$ ($k>0$). 
We employ a Markov-type approximation and assume that  the 
atom reaches its steady state on a timescale that is 
fast as compared to the typical evolution time of $\rft$ 
induced by the atom-cavity coupling. In addition, we suppose that 
the cavity decay rates $\kappa_a$, $\kappa_b$ and the Rabi frequencies $\Omega_a$, 
$\Omega_b$ are small as compared to 
the atomic decay rates $\gamma_{ij}$ and hence neglect 
the contribution of the super-operator $\lio$ to $\mc{L}_0$. 
Under these conditions, the set of equations~(\ref{rho_0}) and~(\ref{iterative}) 
can be solved in a straightforward manner if $\mc{L}_0$ is represented 
by a matrix. However, the procedure is tedious since $H_{\text{C}}$ contains 
the operators $a$ and $b$, and hence one has to keep track of 
the operator ordering. 
Up to third order, 
$\tilde{\vro}\approx \tilde{\vro}^{(0)}+\tilde{\vro}^{(1)}+\tilde{\vro}^{(2)}+\tilde{\vro}^{(3)}$, 
and for vanishing two-photon detuning $\ve=0$ we find
\begin{align}
\tilde{\vro}_{31}  =  \mc{A} g_a |g_b|^2  b^{\dagger} b a \rft , \label{coherences1} \\[0.2cm]
\tilde{\vro}_{42}  = \mc{A} g_b |g_a|^2 b a \rft a^{\dagger},  \label{coherences2}
\end{align}
where 
\be 
\mc{A} = \frac{-1}{(\Delta + i \gamma_{42}/2)| \Omega_{\text{L}}|^2} .
\ee
Finally,  we substitute Eqs.~(\ref{coherences1}) and~(\ref{coherences2}) in Eq.~(\ref{master_field_W}) 
and obtain the master equation~(\ref{master_field}). 
\section{Finite two-photon detuning $\ve\not=0$ \label{sec_cond}}
Here  we establish 
conditions that ensure the validity of Eq.~(\ref{master_field}) for 
non-zero two photon detuning. For simplicity we limit the discussion to 
the case of a purely dissipative photon-photon interaction ($U=\Delta=0$). 
In the ideal case  $\ve=0$, only the third-order term contributes to the coherences in Eqs.~(\ref{coherences1}) 
and~(\ref{coherences2}).  
The first-order term for $\ve\not=0$ gives rise to an additional term $ \mc{L}_1\rf$ on 
the right-hand side of Eq.~(\ref{master_field}). Up to second order in $\ve$, we find 
\begin{align}
 \mc{L}_1\rf = &-\frac{i}{\hbar} [H_1,\rf] \notag \\
&-\frac{\Gamma_1}{2}(a^{\dagger}a\rf+\rf a^{\dagger} a-2a \rf a^{\dagger}),
\end{align}
where 
\begin{align}
& H_1 =-\hbar |g_a|^2 \frac{\delta \ve^2 +\ve|\Omega_{\text{L}}|^2}{2|\Omega_{\text{L}}|^4} a^{\dagger} a \\
& \Gamma_1 = \gamma_{3} \frac{|g_a|^2}{|\Omega_{\text{L}}|^4} \ve^2 ,
\end{align}
and $\gamma_{3} = \gamma_{31} + \gamma_{32}$ is the full decay rate of state $\ket{3}$. 
Here $H_1$ is just a frequency shift of mode $a$, and the term proportional to $\Gamma_1$ 
describes an additional decay channel for photons in mode $a$. This term can be neglected 
provided that 
\be
\Gamma_1= \gamma_{3} \frac{|g_a|^2}{|\Omega_{\text{L}}|^4} \ve^2 \ll \kappa_a. 
\ee
The second-order contribution $\tilde{\vro}^{(2)}$ is equal to zero for all parameters. 
On the other hand, other third-order terms occur for $\ve\not=0$ that give rise to 
other two-particle processes in Eq.~(\ref{master_field}). The magnitude of these 
terms relative to the terms proportional to  $\Gamma$ in Eq.~(\ref{master_field}) 
is negligible provided that $ |\ve|,|\delta| \ll \gamma_{31},\gamma_{32},\gamma_{42}$ and 
\begin{align}
& \frac{|\ve| (\gamma_{3} + \gamma_{42})}{ |\Omega_{\text{L}}|^2} \ll 1 ,
&& \frac{|\delta \ve|}{|\Omega_{\text{L}}|^2}\ll 1.
\end{align}
\section{Definition and measurement of squeezing spectra \label{squeeze}}
The two-mode squeezing spectra are defined as 
\begin{align}
 & S_u(\omega) =2 \int\limits_0^{\infty}\text{d}\tau \cos\omega\tau 
\mean{\mc{T}:\delta \hat{u}_{\text{out}}(t+\tau)\delta \hat{u}_{\text{out}}(t):}, \\
 & S_v(\omega) =2 \int\limits_0^{\infty}\text{d}\tau \cos\omega\tau 
\mean{\mc{T}:\delta \hat{v}_{\text{out}}(t+\tau)\delta\hat{v}_{\text{out}}(t):}, 
\label{susv}
\end{align}
where 
\be
\hat{u}_{\text{out}} = \hat{x}_A + \hat{x}_B ,\qquad 
\hat{v}_{\text{out}} = \hat{p}_A - \hat{p}_B ,
\ee
and the quadratures of the output field are given by
\begin{align}
&\hat{x}_A = \frac{1}{\sqrt{2}}(\aouts e^{-i \phi} + \aouts^{\dagger}e^{i \phi}) ,  \\
\label{quadratures3A}
&\hat{p}_A = \frac{1}{\sqrt{2} i} (\aouts e^{-i \phi} - \aouts^{\dagger}e^{i \phi}) \\
&\hat{x}_B = \frac{1}{\sqrt{2}}(\bouts e^{-i \phi} + \bouts^{\dagger}e^{i \phi}) ,  \\
&\hat{p}_B = \frac{1}{\sqrt{2} i} (\bouts e^{-i \phi} - \bouts^{\dagger}e^{i \phi}).
\label{quadratures3D}
\end{align}
The operator $\mc{T}$ in Eq.~(\ref{susv}) orders products of annihilation operators 
such that their time arguments increase from right to left, and products of 
creation operators are ordered such that time arguments increase from left to right. 
Since the definition of $S_u$ and $S_v$ comprises only expectation values of 
normally and time-ordered products of the output fields, we can easily 
represent $S_u$ and $S_v$ in terms of the cavity fields~\cite{gardiner:85},
\begin{align}
 & S_u(\omega) = 2 \kappa \int\limits_0^{\infty}\text{d}\tau \cos\omega\tau 
\mean{\mc{T}:\delta \hat{u}_{\text{cav}}(t+\tau)\delta \hat{u}_{\text{cav}}(t):}, 
\label{SuvCavityA}\\
 & S_v(\omega) =2 \kappa \int\limits_0^{\infty}\text{d}\tau \cos\omega\tau 
\mean{\mc{T}:\delta \hat{v}_{\text{cav}}(t+\tau)\delta\hat{v}_{\text{cav}}(t):} ,
\label{SuvCavityB}
\end{align}
where $\hat{u}_{\text{cav}}$ and $\hat{v}_{\text{cav}}$ are defined in Eq.~(\ref{u_and_v}). 
%
%
\begin{figure}[t!]
\includegraphics[scale=1]{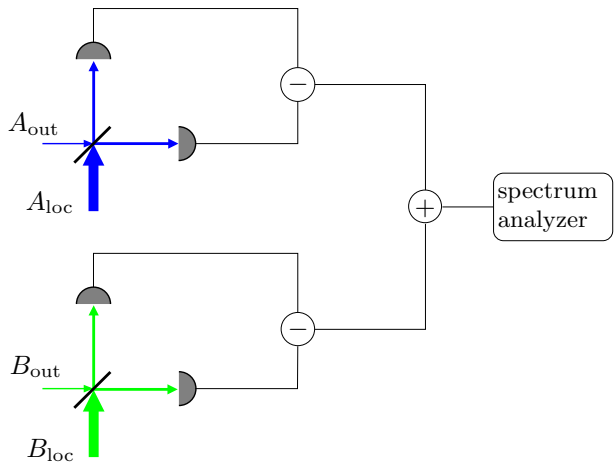}
\caption{\label{picture4} \small (Color online) 
Measurement of the two-mode squeezing spectrum $S_u(\omega)$ in Eq.~(\ref{SuvCavityA}). 
The output field $\aout$ ($\bout$) is superimposed with a strong coherent field 
at a $50:50$ beamsplitter, and the photocurrents of each detector are subtracted. 
The photocurrents of the two  balanced homodyne detections are added and 
fed into a spectrum analyser. The resulting spectrum is given by Eq.~(\ref{pu}) and 
directly proportional to $S_u(\omega)$. 
}
\end{figure}
%
The general setup for the measurement of the two-mode squeezing spectra~\cite{morigi:06,morigi2:06} 
is shown in Fig.~\ref{picture4}. 
The output field $\aout$ ($\bout$) is superimposed with a strong coherent field 
at a $50:50$ beamsplitter, and the photocurrents of each detector are subtracted. 
A similar measurement is performed for the output field $\bout$. The local 
oscillator fields are given by $A_{\text{loc}}=|\alpha| e^{i \theta}$ and 
$B_{\text{loc}}=|\beta| e^{i \theta}$, where $|\alpha|=|\beta|$ and the phase 
$\theta  = \phi-\pi/2$ determines the phase $\phi$ occurring in the 
definition~(\ref{quadratures3A})-(\ref{quadratures3D}) of the quadrature operators. 
The photocurrents of the two balanced homodyne detections  are added and 
the resulting current $i$ is fed into a spectrum analyser. 
The recorded spectrum is then given by~\cite{mandel:qo}
\begin{align}
 P_{u}(\omega) = &\frac{1}{\pi}\int\limits_{0}^{\infty}\text{d}\tau \cos\omega\tau\lim\limits_{t\rightarrow\infty} 
[\mean{i(t)i(t+\tau)} - \mean{i(t)}^2] \notag\\
& = P_{\text{shot}}[1 +  \eta S_u(\omega)], 
\label{pu}
\end{align}
where $P_{\text{shot}}$ is the shot-noise limit and   $\eta$ is the detection efficiency. 
The spectrum $S_v(\omega)$ can be measured with the same setup if the 
phases of the local oscillators are changed according to $\theta \rightarrow \theta+\pi/2$, 
and the photocurrents of the two homodyne detections have to be subtracted rather than added. 
\section{Parameter choices \label{para}}
The validity of our approach requires that the conditions in  Eqs.~(\ref{adcon}) and~(\ref{cond_x}) 
are fulfilled. Here we discuss the possible choices for the system parameters that 
comply with these conditions. 
For simplicity, we assume that the coupling constants, cavity 
decay rates, Rabi frequencies and  the  Fock state cutoffs are identical  for both modes. 
We thus set $|g_a|=|g_b|=|g|$, $\kappa_1=\kappa_2=\kappa$, $\Omega_a=\Omega_b=\Omega$ 
and $N_a=N_b=N$. 
Furthermore, all atomic decay rates are assumed to be the same, 
$\gamma=\gamma_{31}=\gamma_{32}=\gamma_{42}$, and we introduce 
dimensionless parameters $\tilde{p}=p/\gamma$ that are denoted  by a tilde. 
Since we are only interested in the purely dissipative situation, 
we limit the discussion to the case $\Delta=0$ such that the conservative photon-photon interaction $U$ 
vanishes. 
We begin with the cutoff $N$ in the Fock state basis that is  determined by the 
Rabi frequency $\tilde{\Omega}$ driving the cavity modes, the cavity decay rate $\tilde{\kappa}$ 
and the two-photon decay rate $\tilde{\Gamma}$. 
Note that Eq.~(\ref{adcon}) requires these parameters to be much smaller than unity, and 
thus $N$ cannot be arbitrarily large for realistic values of $\tilde{\kappa}$. 
Next we discuss the condition for the two-photon decay rate $\tilde{\Gamma}$ in Eq.~(\ref{adcon}) 
and the inequality in Eq.~(\ref{cond_x}). 
For a given cutoff $N$ of the  Fock state basis, these two conditions read
\be
N\tilde{\Gamma}\ll 1,\quad x= \frac{N |\tilde{g}|^2}{|\tilde{\Omega}_{\text{L}}|^2}   \ll 1. 
\ee 
The two small parameters $x$ and $N\tilde{\Gamma}$ determine the absolute values of the 
coupling constant $\tilde{g}$ and of the Rabi frequency 
$\tilde{\Omega}_{\text{L}}$ corresponding to  the external laser that drives the ion. 
With the explicit  expression for $\tilde{\Gamma}$ in Eq.~(\ref{Gamma}), we find 
\be
|\tilde{g}| = \frac{1}{2} \sqrt{\frac{N \tilde{\Gamma}}{ x}},\quad
|\tilde{\Omega}_{\text{L}}| =  \frac{N\sqrt{\tilde{\Gamma}}}{2x}.
\label{condfin}
\ee
Since $x$ and $N\tilde{\Gamma}$ are both small parameters, one can achieve $|\tilde{g}|<1$ and therefore 
the  strong coupling regime of cavity QED is not required. This is advantageous because 
Eq.~(\ref{condfin}) implies that $|\tilde{\Omega}_{\text{L}}/\tilde{g}|= \sqrt{N/x}$, and 
hence $|\tilde{\Omega}_{\text{L}}|$ can be much larger than  $|\tilde{g}|$. 
It follows that small values of $|\tilde{g}|$ ensure that the scaled Rabi frequency 
$|\tilde{\Omega}_{\text{L}}|$ remains reasonably small.
In addition, a smaller coupling constant facilitates 
the realisation of cavity decay rates that are much smaller than the atomic 
decay rates (see Sec.~\ref{realisation}).
%

%

\begin{thebibliography}{10}%
\makeatletter
\providecommand \@ifxundefined [1]{%
 \ifx #1\undefined \expandafter \@firstoftwo
 \else \expandafter \@secondoftwo
\fi
}%
\providecommand \@ifnum [1]{%
 \ifnum #1\expandafter \@firstoftwo
 \else \expandafter \@secondoftwo
\fi
}%
\providecommand \enquote [1]{``#1''}%
\providecommand \bibnamefont  [1]{#1}%
\providecommand \bibfnamefont [1]{#1}%
\providecommand \citenamefont [1]{#1}%
\providecommand\href[0]{\@sanitize\@href}%
\providecommand\@href[1]{\endgroup\@@startlink{#1}\endgroup\@@href}%
\providecommand\@@href[1]{#1\@@endlink}%
\providecommand \@sanitize [0]{\begingroup\catcode`\&12\catcode`\#12\relax}%
\@ifxundefined \pdfoutput {\@firstoftwo}{%
 \@ifnum{\z@=\pdfoutput}{\@firstoftwo}{\@secondoftwo}%
}{%
 \providecommand\@@startlink[1]{\leavevmode}%
 \providecommand\@@endlink[0]{}%
}{%
 \providecommand\@@startlink[1]{%
  \leavevmode
  \pdfstartlink
   attr{/Border[0 0 1 ]/H/I/C[0 1 1]}%
   user{/Subtype/Link/A<</Type/Action/S/URI/URI(#1)>>}%
  \relax
 }%
 \providecommand\@@endlink[0]{\pdfendlink}%
}%
\providecommand \url  [0]{\begingroup\@sanitize \@url }%
\providecommand \@url [1]{\endgroup\@href {#1}{\urlprefix}}%
\providecommand \urlprefix [0]{URL }%
\providecommand \Eprint[0]{\href }%
\@ifxundefined \urlstyle {%
  \providecommand \doi [1]{doi:\discretionary{}{}{}#1}%
}{%
  \providecommand \doi [0]{doi:\discretionary{}{}{}\begingroup
  \urlstyle{rm}\Url }%
}%
\providecommand \doibase [0]{http://dx.doi.org/}%
\providecommand \Doi[1]{\href{\doibase#1}}%
\providecommand \bibAnnote [3]{%
  \BibitemShut{#1}%
  \begin{quotation}\noindent
    \textsc{Key:}\ #2\\\textsc{Annotation:}\ #3%
  \end{quotation}%
}%
\providecommand \bibAnnoteFile [2]{%
  \IfFileExists{#2}{\bibAnnote {#1} {#2} {\input{#2}}}{}%
}%
\providecommand \typeout [0]{\immediate \write \m@ne }%
\providecommand \selectlanguage [0]{\@gobble}%
\providecommand \bibinfo [0]{\@secondoftwo}%
\providecommand \bibfield [0]{\@secondoftwo}%
\providecommand \translation [1]{[#1]}%
\providecommand \BibitemOpen[0]{}%
\providecommand \bibitemStop [0]{}%
\providecommand \bibitemNoStop [0]{.\EOS\space}%
\providecommand \EOS [0]{\spacefactor3000\relax}%
\providecommand \BibitemShut [1]{\csname bibitem#1\endcsname}%
\bibitem{braunstein:05}%
  \BibitemOpen
  \bibfield{author}{%
  \bibinfo {author} {\bibfnamefont{S.~L.}\ \bibnamefont{Braunstein}}\ and\
  \bibinfo {author} {\bibfnamefont{P.}~\bibnamefont{van Look}},\ }%
  \bibfield{journal}{%
  \bibinfo {journal} {Rev. Mod. Phys.}\ }%
  \textbf{\bibinfo {volume} {77}},\ \bibinfo {pages} {513} (\bibinfo {year}
  {2005})%
  \bibAnnoteFile{NoStop}{braunstein:05}%
\bibitem{dorner:09}%
  \BibitemOpen
  \bibfield{author}{%
  \bibinfo {author} {\bibfnamefont{U.}~\bibnamefont{Dorner}}, \bibinfo {author}
  {\bibfnamefont{R.}~\bibnamefont{Demkowicz-Dobrzanski}}, \bibinfo {author}
  {\bibfnamefont{B.~J.}\ \bibnamefont{Smith}}, \bibinfo {author}
  {\bibfnamefont{J.~S.}\ \bibnamefont{Lundeen}}, \bibinfo {author}
  {\bibfnamefont{W.}~\bibnamefont{Wasilewski}}, \bibinfo {author}
  {\bibfnamefont{K.}~\bibnamefont{Banaszek}},\ and\ \bibinfo {author}
  {\bibfnamefont{I.~A.}\ \bibnamefont{Walmsley}},\ }%
  \bibfield{journal}{%
  \bibinfo {journal} {Phys. Rev. Lett.}\ }%
  \textbf{\bibinfo {volume} {102}},\ \bibinfo {pages} {040403} (\bibinfo {year}
  {2009})%
  \bibAnnoteFile{NoStop}{dorner:09}%
\bibitem{boto:00}%
  \BibitemOpen
  \bibfield{author}{%
  \bibinfo {author} {\bibfnamefont{A.~N.}\ \bibnamefont{Boto}}, \bibinfo
  {author} {\bibfnamefont{P.}~\bibnamefont{Kok}}, \bibinfo {author}
  {\bibfnamefont{D.~S.}\ \bibnamefont{Abrams}}, \bibinfo {author}
  {\bibfnamefont{S.~L.}\ \bibnamefont{Braunstein}}, \bibinfo {author}
  {\bibfnamefont{C.~P.}\ \bibnamefont{Williams}},\ and\ \bibinfo {author}
  {\bibfnamefont{J.~P.}\ \bibnamefont{Dowling}},\ }%
  \bibfield{journal}{%
  \bibinfo {journal} {Phys. Rev. Lett.}\ }%
  \textbf{\bibinfo {volume} {85}},\ \bibinfo {pages} {2733} (\bibinfo {year}
  {2000})%
  \bibAnnoteFile{NoStop}{boto:00}%
\bibitem{bollinger:96}%
  \BibitemOpen
  \bibfield{author}{%
  \bibinfo {author} {\bibfnamefont{J.~J.}\ \bibnamefont{Bollinger}}, \bibinfo
  {author} {\bibfnamefont{W.~M.}\ \bibnamefont{Itano}}, \bibinfo {author}
  {\bibfnamefont{D.~J.}\ \bibnamefont{Wineland}},\ and\ \bibinfo {author}
  {\bibfnamefont{D.~J.}\ \bibnamefont{Heinzen}},\ }%
  \bibfield{journal}{%
  \bibinfo {journal} {Phys. Rev. A}\ }%
  \textbf{\bibinfo {volume} {54}},\ \bibinfo {pages} {R4649} (\bibinfo {year}
  {1996})%
  \bibAnnoteFile{NoStop}{bollinger:96}%
\bibitem{dowling:08}%
  \BibitemOpen
  \bibfield{author}{%
  \bibinfo {author} {\bibfnamefont{J.~P.}\ \bibnamefont{Dowling}},\ }%
  \bibfield{journal}{%
  \bibinfo {journal} {Contemp. Phys.}\ }%
  \textbf{\bibinfo {volume} {49}},\ \bibinfo {pages} {125} (\bibinfo {year}
  {2008})%
  \bibAnnoteFile{NoStop}{dowling:08}%
\bibitem{kacprowicz:10}%
  \BibitemOpen
  \bibfield{author}{%
  \bibinfo {author} {\bibfnamefont{M.}~\bibnamefont{Kacprowicz}}, \bibinfo
  {author} {\bibfnamefont{R.}~\bibnamefont{Demkowicz-Dobrza\'{n}ski}}, \bibinfo
  {author} {\bibfnamefont{W.}~\bibnamefont{Wasilewski}}, \bibinfo {author}
  {\bibfnamefont{K.}~\bibnamefont{Banaszek}},\ and\ \bibinfo {author}
  {\bibfnamefont{I.~A.}\ \bibnamefont{Walmsley}},\ }%
  \bibfield{journal}{%
  \bibinfo {journal} {Nature Photonics}\ }%
  \textbf{\bibinfo {volume} {4}},\ \bibinfo {pages} {357} (\bibinfo {year}
  {2010})%
  \bibAnnoteFile{NoStop}{kacprowicz:10}%
\bibitem{escher:11}%
  \BibitemOpen
  \bibfield{author}{%
  \bibinfo {author} {\bibfnamefont{B.~M.}\ \bibnamefont{Escher}}, \bibinfo
  {author} {\bibfnamefont{R.~L.}\ \bibnamefont{de~Matos~Filho}},\ and\ \bibinfo
  {author} {\bibfnamefont{L.}~\bibnamefont{Davidovich}},\ }%
  \bibfield{journal}{%
  \bibinfo {journal} {Nat. Phys.}\ }%
  \textbf{\bibinfo {volume} {7}},\ \bibinfo {pages} {406} (\bibinfo {year}
  {2011})%
  \bibAnnoteFile{NoStop}{escher:11}%
\bibitem{thomaspeter:11}%
  \BibitemOpen
  \bibfield{author}{%
  \bibinfo {author} {\bibfnamefont{N.}~\bibnamefont{Thomas-Peter}}, \bibinfo
  {author} {\bibfnamefont{B.~J.}\ \bibnamefont{Smith}}, \bibinfo {author}
  {\bibfnamefont{A.}~\bibnamefont{Datta}}, \bibinfo {author}
  {\bibfnamefont{L.}~\bibnamefont{Zhang}}, \bibinfo {author}
  {\bibfnamefont{U.}~\bibnamefont{Dorner}},\ and\ \bibinfo {author}
  {\bibfnamefont{I.~A.}\ \bibnamefont{Walmsley}},\ }%
  \bibfield{journal}{%
  \bibinfo {journal} {Phys. Rev. Lett.}\ }%
  \textbf{\bibinfo {volume} {107}},\ \bibinfo {pages} {113603} (\bibinfo {year}
  {2011})%
  \bibAnnoteFile{NoStop}{thomaspeter:11}%
\bibitem{sanders:92}%
  \BibitemOpen
  \bibfield{author}{%
  \bibinfo {author} {\bibfnamefont{B.~C.}\ \bibnamefont{Sanders}},\ }%
  \bibfield{journal}{%
  \bibinfo {journal} {Phys. Rev. A}\ }%
  \textbf{\bibinfo {volume} {45}},\ \bibinfo {pages} {6811} (\bibinfo {year}
  {1992})%
  \bibAnnoteFile{NoStop}{sanders:92}%
\bibitem{sanders:11}%
  \BibitemOpen
  \bibinfo {note} {B. C. Sanders, arXiv:1112.1778v1.}%
  \bibAnnoteFile{Stop}{sanders:11}%
\bibitem{gerry:01}%
  \BibitemOpen
  \bibfield{author}{%
  \bibinfo {author} {\bibfnamefont{C.~C.}\ \bibnamefont{Gerry}}\ and\ \bibinfo
  {author} {\bibfnamefont{R.~A.}\ \bibnamefont{Campos}},\ }%
  \bibfield{journal}{%
  \bibinfo {journal} {Phys. Rev. A}\ }%
  \textbf{\bibinfo {volume} {64}},\ \bibinfo {pages} {063814} (\bibinfo {year}
  {2001})%
  \bibAnnoteFile{NoStop}{gerry:01}%
\bibitem{gerry:10}%
  \BibitemOpen
  \bibfield{author}{%
  \bibinfo {author} {\bibfnamefont{C.~C.}\ \bibnamefont{Gerry}}\ and\ \bibinfo
  {author} {\bibfnamefont{J.}~\bibnamefont{Mimih}},\ }%
  \bibfield{journal}{%
  \bibinfo {journal} {Contemporary Physics}\ }%
  \textbf{\bibinfo {volume} {51}},\ \bibinfo {pages} {497} (\bibinfo {year}
  {2010})%
  \bibAnnoteFile{NoStop}{gerry:10}%
\bibitem{joo:11}%
  \BibitemOpen
  \bibfield{author}{%
  \bibinfo {author} {\bibfnamefont{J.}~\bibnamefont{Joo}}, \bibinfo {author}
  {\bibfnamefont{W.~J.}\ \bibnamefont{Munro}},\ and\ \bibinfo {author}
  {\bibfnamefont{T.~P.}\ \bibnamefont{Spiller}},\ }%
  \bibfield{journal}{%
  \bibinfo {journal} {Phys. Rev. Lett.}\ }%
  \textbf{\bibinfo {volume} {107}},\ \bibinfo {pages} {083601} (\bibinfo {year}
  {2011})%
  \bibAnnoteFile{NoStop}{joo:11}%
\bibitem{mitchell:04}%
  \BibitemOpen
  \bibfield{author}{%
  \bibinfo {author} {\bibfnamefont{M.~W.}\ \bibnamefont{Mitchell}}, \bibinfo
  {author} {\bibfnamefont{J.~S.}\ \bibnamefont{Lundeen}},\ and\ \bibinfo
  {author} {\bibfnamefont{A.~M.}\ \bibnamefont{Steinberg}},\ }%
  \bibfield{journal}{%
  \bibinfo {journal} {Nature}\ }%
  \textbf{\bibinfo {volume} {429}},\ \bibinfo {pages} {161} (\bibinfo {year}
  {2004})%
  \bibAnnoteFile{NoStop}{mitchell:04}%
\bibitem{afek:10}%
  \BibitemOpen
  \bibfield{author}{%
  \bibinfo {author} {\bibfnamefont{I.}~\bibnamefont{Afek}}, \bibinfo {author}
  {\bibfnamefont{O.}~\bibnamefont{Ambar}},\ and\ \bibinfo {author}
  {\bibfnamefont{Y.}~\bibnamefont{Silberberg}},\ }%
  \bibfield{journal}{%
  \bibinfo {journal} {Science}\ }%
  \textbf{\bibinfo {volume} {328}},\ \bibinfo {pages} {879} (\bibinfo {year}
  {2010})%
  \bibAnnoteFile{NoStop}{afek:10}%
\bibitem{ourjoumtsev:06}%
  \BibitemOpen
  \bibfield{author}{%
  \bibinfo {author} {\bibfnamefont{A.}~\bibnamefont{Ourjoumtsev}}, \bibinfo
  {author} {\bibfnamefont{R.}~\bibnamefont{Tualle-Brouri}}, \bibinfo {author}
  {\bibfnamefont{J.}~\bibnamefont{Laurat}},\ and\ \bibinfo {author}
  {\bibfnamefont{P.}~\bibnamefont{Grangier}},\ }%
  \bibfield{journal}{%
  \bibinfo {journal} {Science}\ }%
  \textbf{\bibinfo {volume} {312}},\ \bibinfo {pages} {83} (\bibinfo {year}
  {2006})%
  \bibAnnoteFile{NoStop}{ourjoumtsev:06}%
\bibitem{ourjoumtsev:07}%
  \BibitemOpen
  \bibfield{author}{%
  \bibinfo {author} {\bibfnamefont{A.}~\bibnamefont{Ourjoumtsev}}, \bibinfo
  {author} {\bibfnamefont{H.}~\bibnamefont{Jeong}}, \bibinfo {author}
  {\bibfnamefont{R.}~\bibnamefont{Tualle-Brouri}},\ and\ \bibinfo {author}
  {\bibfnamefont{P.}~\bibnamefont{Grangier}},\ }%
  \bibfield{journal}{%
  \bibinfo {journal} {Nature}\ }%
  \textbf{\bibinfo {volume} {448}},\ \bibinfo {pages} {784} (\bibinfo {year}
  {2007})%
  \bibAnnoteFile{NoStop}{ourjoumtsev:07}%
\bibitem{kraus:08}%
  \BibitemOpen
  \bibfield{author}{%
  \bibinfo {author} {\bibfnamefont{B.}~\bibnamefont{Kraus}}, \bibinfo {author}
  {\bibfnamefont{H.~P.}\ \bibnamefont{B\"uchler}}, \bibinfo {author}
  {\bibfnamefont{S.}~\bibnamefont{Diehl}}, \bibinfo {author}
  {\bibfnamefont{A.}~\bibnamefont{Kantian}}, \bibinfo {author}
  {\bibfnamefont{A.}~\bibnamefont{Micheli}},\ and\ \bibinfo {author}
  {\bibfnamefont{P.}~\bibnamefont{Zoller}},\ }%
  \bibfield{journal}{%
  \bibinfo {journal} {Phys. Rev. A}\ }%
  \textbf{\bibinfo {volume} {78}},\ \bibinfo {pages} {042307} (\bibinfo {year}
  {2008})%
  \bibAnnoteFile{NoStop}{kraus:08}%
\bibitem{verstraete:09}%
  \BibitemOpen
  \bibfield{author}{%
  \bibinfo {author} {\bibfnamefont{F.}~\bibnamefont{Verstraete}}, \bibinfo
  {author} {\bibfnamefont{M.~M.}\ \bibnamefont{Wolf}},\ and\ \bibinfo {author}
  {\bibfnamefont{J.~I.}\ \bibnamefont{Cirac}},\ }%
  \bibfield{journal}{%
  \bibinfo {journal} {Nat. Phys.}\ }%
  \textbf{\bibinfo {volume} {5}},\ \bibinfo {pages} {633} (\bibinfo {year}
  {2009})%
  \bibAnnoteFile{NoStop}{verstraete:09}%
\bibitem{diehl:08}%
  \BibitemOpen
  \bibfield{author}{%
  \bibinfo {author} {\bibfnamefont{S.}~\bibnamefont{Diehl}}, \bibinfo {author}
  {\bibfnamefont{A.}~\bibnamefont{Micheli}}, \bibinfo {author}
  {\bibfnamefont{A.}~\bibnamefont{Kantian}}, \bibinfo {author}
  {\bibfnamefont{B.}~\bibnamefont{Kraus}}, \bibinfo {author}
  {\bibfnamefont{H.~P.}\ \bibnamefont{B\"uchler}},\ and\ \bibinfo {author}
  {\bibfnamefont{P.}~\bibnamefont{Zoller}},\ }%
  \bibfield{journal}{%
  \bibinfo {journal} {Nat. Phys.}\ }%
  \textbf{\bibinfo {volume} {4}},\ \bibinfo {pages} {878} (\bibinfo {year}
  {2008})%
  \bibAnnoteFile{NoStop}{diehl:08}%
\bibitem{syassen:08}%
  \BibitemOpen
  \bibfield{author}{%
  \bibinfo {author} {\bibfnamefont{N.}~\bibnamefont{Syassen}}, \bibinfo
  {author} {\bibfnamefont{D.~M.}\ \bibnamefont{Bauer}}, \bibinfo {author}
  {\bibfnamefont{M.}~\bibnamefont{Lettner}}, \bibinfo {author}
  {\bibfnamefont{T.}~\bibnamefont{Volz}}, \bibinfo {author}
  {\bibfnamefont{D.}~\bibnamefont{Dietze}}, \bibinfo {author}
  {\bibfnamefont{J.~J.}\ \bibnamefont{Garc\'ia-Ripoll}}, \bibinfo {author}
  {\bibfnamefont{J.~I.}\ \bibnamefont{Cirac}}, \bibinfo {author}
  {\bibfnamefont{G.}~\bibnamefont{Rempe}},\ and\ \bibinfo {author}
  {\bibfnamefont{S.}~\bibnamefont{D\"urr}},\ }%
  \bibfield{journal}{%
  \bibinfo {journal} {Science}\ }%
  \textbf{\bibinfo {volume} {320}},\ \bibinfo {pages} {1329} (\bibinfo {year}
  {2008})%
  \bibAnnoteFile{NoStop}{syassen:08}%
\bibitem{ripoll:08}%
  \BibitemOpen
  \bibfield{author}{%
  \bibinfo {author} {\bibfnamefont{J.~J.}\ \bibnamefont{Garc\'ia-Ripoll}},
  \bibinfo {author} {\bibfnamefont{S.}~\bibnamefont{D\"urr}}, \bibinfo {author}
  {\bibfnamefont{N.}~\bibnamefont{Syassen}}, \bibinfo {author}
  {\bibfnamefont{D.~M.}\ \bibnamefont{Bauer}}, \bibinfo {author}
  {\bibfnamefont{M.}~\bibnamefont{Lettner}}, \bibinfo {author}
  {\bibfnamefont{G.}~\bibnamefont{Rempe}},\ and\ \bibinfo {author}
  {\bibfnamefont{J.~I.}\ \bibnamefont{Cirac}},\ }%
  \bibfield{journal}{%
  \bibinfo {journal} {New. J. Phys.}\ }%
  \textbf{\bibinfo {volume} {11}},\ \bibinfo {pages} {013053} (\bibinfo {year}
  {2009})%
  \bibAnnoteFile{NoStop}{ripoll:08}%
\bibitem{duerr:09}%
  \BibitemOpen
  \bibfield{author}{%
  \bibinfo {author} {\bibfnamefont{S.}~\bibnamefont{D\"urr}}, \bibinfo {author}
  {\bibfnamefont{J.~J.}\ \bibnamefont{Garc\'ia-Ripoll}}, \bibinfo {author}
  {\bibfnamefont{N.}~\bibnamefont{Syassen}}, \bibinfo {author}
  {\bibfnamefont{D.~M.}\ \bibnamefont{Bauer}}, \bibinfo {author}
  {\bibfnamefont{M.}~\bibnamefont{Lettner}}, \bibinfo {author}
  {\bibfnamefont{J.~I.}\ \bibnamefont{Cirac}},\ and\ \bibinfo {author}
  {\bibfnamefont{G.}~\bibnamefont{Rempe}},\ }%
  \bibfield{journal}{%
  \bibinfo {journal} {Phys. Rev. A}\ }%
  \textbf{\bibinfo {volume} {79}},\ \bibinfo {pages} {023614} (\bibinfo {year}
  {2009})%
  \bibAnnoteFile{NoStop}{duerr:09}%
\bibitem{daley:09}%
  \BibitemOpen
  \bibfield{author}{%
  \bibinfo {author} {\bibfnamefont{A.~J.}\ \bibnamefont{Daley}}, \bibinfo
  {author} {\bibfnamefont{J.~M.}\ \bibnamefont{Taylor}}, \bibinfo {author}
  {\bibfnamefont{S.}~\bibnamefont{Diehl}}, \bibinfo {author}
  {\bibfnamefont{M.}~\bibnamefont{Baranov}},\ and\ \bibinfo {author}
  {\bibfnamefont{P.}~\bibnamefont{Zoller}},\ }%
  \bibfield{journal}{%
  \bibinfo {journal} {Phys. Rev. Lett.}\ }%
  \textbf{\bibinfo {volume} {102}},\ \bibinfo {pages} {040402} (\bibinfo {year}
  {2009})%
  \bibAnnoteFile{NoStop}{daley:09}%
\bibitem{kiffner:10}%
  \BibitemOpen
  \bibfield{author}{%
  \bibinfo {author} {\bibfnamefont{M.}~\bibnamefont{Kiffner}}\ and\ \bibinfo
  {author} {\bibfnamefont{M.~J.}\ \bibnamefont{Hartmann}},\ }%
  \bibfield{journal}{%
  \bibinfo {journal} {Phys. Rev. A}\ }%
  \textbf{\bibinfo {volume} {81}},\ \bibinfo {pages} {021806(R)} (\bibinfo
  {year} {2010})%
  \bibAnnoteFile{NoStop}{kiffner:10}%
\bibitem{kiffner:10b}%
  \BibitemOpen
  \bibfield{author}{%
  \bibinfo {author} {\bibfnamefont{M.}~\bibnamefont{Kiffner}}\ and\ \bibinfo
  {author} {\bibfnamefont{M.~J.}\ \bibnamefont{Hartmann}},\ }%
  \bibfield{journal}{%
  \bibinfo {journal} {Phys. Rev. A}\ }%
  \textbf{\bibinfo {volume} {82}},\ \bibinfo {pages} {033813} (\bibinfo {year}
  {2010})%
  \bibAnnoteFile{NoStop}{kiffner:10b}%
\bibitem{kiffner:11}%
  \BibitemOpen
  \bibfield{author}{%
  \bibinfo {author} {\bibfnamefont{M.}~\bibnamefont{Kiffner}}\ and\ \bibinfo
  {author} {\bibfnamefont{M.~J.}\ \bibnamefont{Hartmann}},\ }%
  \bibfield{journal}{%
  \bibinfo {journal} {New. J. Phys.}\ }%
  \textbf{\bibinfo {volume} {13}},\ \bibinfo {pages} {053027} (\bibinfo {year}
  {2011})%
  \bibAnnoteFile{NoStop}{kiffner:11}%
\bibitem{krauter:10b}%
  \BibitemOpen
  \bibfield{author}{%
  \bibinfo {author} {\bibfnamefont{H.}~\bibnamefont{Krauter}}, \bibinfo
  {author} {\bibfnamefont{C.~A.}\ \bibnamefont{Muschik}}, \bibinfo {author}
  {\bibfnamefont{K.}~\bibnamefont{Jensen}}, \bibinfo {author}
  {\bibfnamefont{W.}~\bibnamefont{Wasilewski}}, \bibinfo {author}
  {\bibfnamefont{J.~M.}\ \bibnamefont{Petersen}}, \bibinfo {author}
  {\bibfnamefont{J.~I.}\ \bibnamefont{Cirac}},\ and\ \bibinfo {author}
  {\bibfnamefont{E.~S.}\ \bibnamefont{Polzik}},\ }%
  \bibfield{journal}{%
  \bibinfo {journal} {Phys. Rev. Lett.}\ }%
  \textbf{\bibinfo {volume} {107}},\ \bibinfo {pages} {080503} (\bibinfo {year}
  {2011})%
  \bibAnnoteFile{NoStop}{krauter:10b}%
\bibitem{muschik:10b}%
  \BibitemOpen
  \bibfield{author}{%
  \bibinfo {author} {\bibfnamefont{C.~A.}\ \bibnamefont{Muschik}}, \bibinfo
  {author} {\bibfnamefont{E.~S.}\ \bibnamefont{Polzik}},\ and\ \bibinfo
  {author} {\bibfnamefont{J.~I.}\ \bibnamefont{Cirac}},\ }%
  \bibfield{journal}{%
  \bibinfo {journal} {Phys. Rev. A}\ }%
  \textbf{\bibinfo {volume} {83}},\ \bibinfo {pages} {052312} (\bibinfo {year}
  {2011})%
  \bibAnnoteFile{NoStop}{muschik:10b}%
\bibitem{barreiro:11}%
  \BibitemOpen
  \bibfield{author}{%
  \bibinfo {author} {\bibfnamefont{J.~T.}\ \bibnamefont{Barreiro}}, \bibinfo
  {author} {\bibfnamefont{M.}~\bibnamefont{M\"uller}}, \bibinfo {author}
  {\bibfnamefont{P.}~\bibnamefont{Schindler}}, \bibinfo {author}
  {\bibfnamefont{D.}~\bibnamefont{Nigg}}, \bibinfo {author}
  {\bibfnamefont{T.}~\bibnamefont{Monz}}, \bibinfo {author}
  {\bibfnamefont{M.}~\bibnamefont{Chwalla}}, \bibinfo {author}
  {\bibfnamefont{M.}~\bibnamefont{Hennrich}}, \bibinfo {author}
  {\bibfnamefont{C.~F.}\ \bibnamefont{Roos}}, \bibinfo {author}
  {\bibfnamefont{P.}~\bibnamefont{Zoller}},\ and\ \bibinfo {author}
  {\bibfnamefont{R.}~\bibnamefont{Blatt}},\ }%
  \bibfield{journal}{%
  \bibinfo {journal} {Nature}\ }%
  \textbf{\bibinfo {volume} {470}},\ \bibinfo {pages} {486} (\bibinfo {year}
  {2011})%
  \bibAnnoteFile{NoStop}{barreiro:11}%
\bibitem{mueller:11}%
  \BibitemOpen
  \bibinfo {note} {M. M\"uller, K. Hammerer, Y. L. Zhou, C. F. Roos, and P.
  Zoller, arXiv:1104.2507v1.}%
  \bibAnnoteFile{Stop}{mueller:11}%
\bibitem{vidal:02}%
  \BibitemOpen
  \bibfield{author}{%
  \bibinfo {author} {\bibfnamefont{G.}~\bibnamefont{Vidal}}\ and\ \bibinfo
  {author} {\bibfnamefont{R.~F.}\ \bibnamefont{Werner}},\ }%
  \bibfield{journal}{%
  \bibinfo {journal} {Phys. Rev. A}\ }%
  \textbf{\bibinfo {volume} {65}},\ \bibinfo {pages} {032314} (\bibinfo {year}
  {2002})%
  \bibAnnoteFile{NoStop}{vidal:02}%
\bibitem{duan:00}%
  \BibitemOpen
  \bibfield{author}{%
  \bibinfo {author} {\bibfnamefont{L.-M.}\ \bibnamefont{Duan}}, \bibinfo
  {author} {\bibfnamefont{G.}~\bibnamefont{Giedke}}, \bibinfo {author}
  {\bibfnamefont{J.~I.}\ \bibnamefont{Cirac}},\ and\ \bibinfo {author}
  {\bibfnamefont{P.}~\bibnamefont{Zoller}},\ }%
  \bibfield{journal}{%
  \bibinfo {journal} {Phys. Rev. Lett.}\ }%
  \textbf{\bibinfo {volume} {84}},\ \bibinfo {pages} {2722} (\bibinfo {year}
  {2000})%
  \bibAnnoteFile{NoStop}{duan:00}%
\bibitem{schmidt:96}%
  \BibitemOpen
  \bibfield{author}{%
  \bibinfo {author} {\bibfnamefont{H.}~\bibnamefont{Schmidt}}\ and\ \bibinfo
  {author} {\bibfnamefont{A.}~\bibnamefont{Imamo\v{g}lu}},\ }%
  \bibfield{journal}{%
  \bibinfo {journal} {Opt. Lett.}\ }%
  \textbf{\bibinfo {volume} {21}},\ \bibinfo {pages} {1936} (\bibinfo {year}
  {1996})%
  \bibAnnoteFile{NoStop}{schmidt:96}%
\bibitem{kang:03}%
  \BibitemOpen
  \bibfield{author}{%
  \bibinfo {author} {\bibfnamefont{H.}~\bibnamefont{Kang}}\ and\ \bibinfo
  {author} {\bibfnamefont{Y.}~\bibnamefont{Zhu}},\ }%
  \bibfield{journal}{%
  \bibinfo {journal} {Phys. Rev. Lett.}\ }%
  \textbf{\bibinfo {volume} {91}},\ \bibinfo {pages} {093601} (\bibinfo {year}
  {2003})%
  \bibAnnoteFile{NoStop}{kang:03}%
\bibitem{harris:98}%
  \BibitemOpen
  \bibfield{author}{%
  \bibinfo {author} {\bibfnamefont{S.~E.}\ \bibnamefont{Harris}}\ and\ \bibinfo
  {author} {\bibfnamefont{Y.}~\bibnamefont{Yamamoto}},\ }%
  \bibfield{journal}{%
  \bibinfo {journal} {Phys. Rev. Lett.}\ }%
  \textbf{\bibinfo {volume} {81}},\ \bibinfo {pages} {3611} (\bibinfo {year}
  {1998})%
  \bibAnnoteFile{NoStop}{harris:98}%
\bibitem{comment1}%
  \BibitemOpen
  \bibinfo {note} {C. Cohen-Tannoudji, J. Dupont-Roc, G. Grynberg,
  \textit{Atom-Photon Interactions} (Wiley, New York, 1992), Sec. III.C.3.}%
  \bibAnnoteFile{Stop}{comment1}%
\bibitem{gardiner:85}%
  \BibitemOpen
  \bibfield{author}{%
  \bibinfo {author} {\bibfnamefont{C.~W.}\ \bibnamefont{Gardiner}}\ and\
  \bibinfo {author} {\bibfnamefont{M.~J.}\ \bibnamefont{Collett}},\ }%
  \bibfield{journal}{%
  \bibinfo {journal} {Phys. Rev. A}\ }%
  \textbf{\bibinfo {volume} {31}},\ \bibinfo {pages} {3761} (\bibinfo {year}
  {1985})%
  \bibAnnoteFile{NoStop}{gardiner:85}%
\bibitem{clerk:10}%
  \BibitemOpen
  \bibfield{author}{%
  \bibinfo {author} {\bibfnamefont{A.~A.}\ \bibnamefont{Clerk}}, \bibinfo
  {author} {\bibfnamefont{M.~H.}\ \bibnamefont{Devoret}}, \bibinfo {author}
  {\bibfnamefont{S.~M.}\ \bibnamefont{Girvin}}, \bibinfo {author}
  {\bibfnamefont{F.}~\bibnamefont{Marquardt}},\ and\ \bibinfo {author}
  {\bibfnamefont{R.~J.}\ \bibnamefont{Schoelkopf}},\ }%
  \bibfield{journal}{%
  \bibinfo {journal} {Rev. Mod. Phys.}\ }%
  \textbf{\bibinfo {volume} {82}},\ \bibinfo {pages} {1155} (\bibinfo {year}
  {2010})%
  \bibAnnoteFile{NoStop}{clerk:10}%
\bibitem{agarwal:qst2}%
  \BibitemOpen
  \bibinfo {note} {G. S. Agarwal, in \textit{Quantum Statistical Theories of
  Spontaneous Emission and Their Relation to Other Approaches}, edited by G.
  H\"ohler (Springer, Berlin, 1974).}%
  \bibAnnoteFile{Stop}{agarwal:qst2}%
\bibitem{morigi:06}%
  \BibitemOpen
  \bibfield{author}{%
  \bibinfo {author} {\bibfnamefont{G.}~\bibnamefont{Morigi}}, \bibinfo {author}
  {\bibfnamefont{J.}~\bibnamefont{Eschner}}, \bibinfo {author}
  {\bibfnamefont{S.}~\bibnamefont{Mancini}},\ and\ \bibinfo {author}
  {\bibfnamefont{D.}~\bibnamefont{Vitali}},\ }%
  \bibfield{journal}{%
  \bibinfo {journal} {Phys. Rev. Lett.}\ }%
  \textbf{\bibinfo {volume} {96}},\ \bibinfo {pages} {023601} (\bibinfo {year}
  {2006})%
  \bibAnnoteFile{NoStop}{morigi:06}%
\bibitem{morigi2:06}%
  \BibitemOpen
  \bibfield{author}{%
  \bibinfo {author} {\bibfnamefont{G.}~\bibnamefont{Morigi}}, \bibinfo {author}
  {\bibfnamefont{J.}~\bibnamefont{Eschner}}, \bibinfo {author}
  {\bibfnamefont{S.}~\bibnamefont{Mancini}},\ and\ \bibinfo {author}
  {\bibfnamefont{D.}~\bibnamefont{Vitali}},\ }%
  \bibfield{journal}{%
  \bibinfo {journal} {Phys. Rev. A}\ }%
  \textbf{\bibinfo {volume} {73}},\ \bibinfo {pages} {033822} (\bibinfo {year}
  {2006})%
  \bibAnnoteFile{NoStop}{morigi2:06}%
\bibitem{jing:06}%
  \BibitemOpen
  \bibfield{author}{%
  \bibinfo {author} {\bibfnamefont{J.}~\bibnamefont{Jing}}, \bibinfo {author}
  {\bibfnamefont{S.}~\bibnamefont{Feng}}, \bibinfo {author}
  {\bibfnamefont{R.}~\bibnamefont{Bloomer}},\ and\ \bibinfo {author}
  {\bibfnamefont{O.}~\bibnamefont{Pfister}},\ }%
  \bibfield{journal}{%
  \bibinfo {journal} {Phys. Rev. A}\ }%
  \textbf{\bibinfo {volume} {74}},\ \bibinfo {pages} {041804(R)} (\bibinfo
  {year} {2006})%
  \bibAnnoteFile{NoStop}{jing:06}%
\bibitem{weber:09}%
  \BibitemOpen
  \bibfield{author}{%
  \bibinfo {author} {\bibfnamefont{B.}~\bibnamefont{Weber}}, \bibinfo {author}
  {\bibfnamefont{H.~P.}\ \bibnamefont{Specht}}, \bibinfo {author}
  {\bibfnamefont{T.}~\bibnamefont{M\"uller}}, \bibinfo {author}
  {\bibfnamefont{J.}~\bibnamefont{Bochmann}}, \bibinfo {author}
  {\bibfnamefont{M.}~\bibnamefont{M\"ucke}}, \bibinfo {author}
  {\bibfnamefont{D.~L.}\ \bibnamefont{Moehring}},\ and\ \bibinfo {author}
  {\bibfnamefont{G.}~\bibnamefont{Rempe}},\ }%
  \bibfield{journal}{%
  \bibinfo {journal} {Phys. Rev. Lett.}\ }%
  \textbf{\bibinfo {volume} {102}},\ \bibinfo {pages} {030501} (\bibinfo {year}
  {2009})%
  \bibAnnoteFile{NoStop}{weber:09}%
\bibitem{khudaverdyan:09}%
  \BibitemOpen
  \bibfield{author}{%
  \bibinfo {author} {\bibfnamefont{M.}~\bibnamefont{Khudaverdyan}}, \bibinfo
  {author} {\bibfnamefont{W.}~\bibnamefont{Alt}}, \bibinfo {author}
  {\bibfnamefont{T.}~\bibnamefont{Kampschulte}}, \bibinfo {author}
  {\bibfnamefont{S.}~\bibnamefont{Reick}}, \bibinfo {author}
  {\bibfnamefont{A.}~\bibnamefont{Thobe}}, \bibinfo {author}
  {\bibfnamefont{A.}~\bibnamefont{Widera}},\ and\ \bibinfo {author}
  {\bibfnamefont{D.}~\bibnamefont{Meschede}},\ }%
  \bibfield{journal}{%
  \bibinfo {journal} {Phys. Rev. Lett.}\ }%
  \textbf{\bibinfo {volume} {103}},\ \bibinfo {pages} {123006} (\bibinfo {year}
  {2009})%
  \bibAnnoteFile{NoStop}{khudaverdyan:09}%
\bibitem{stute:11a}%
  \BibitemOpen
  \bibinfo {note} {A. Stute, B. Casabone, B. Brandst\"atter, D. Habicher, P. O.
  Schmidt, T. E. Northup, and R. Blatt, arXv:1105.0579v1.}%
  \bibAnnoteFile{Stop}{stute:11a}%
\bibitem{russo:09}%
  \BibitemOpen
  \bibfield{author}{%
  \bibinfo {author} {\bibfnamefont{C.}~\bibnamefont{Russo}}, \bibinfo {author}
  {\bibfnamefont{H.}~\bibnamefont{Barros}}, \bibinfo {author}
  {\bibfnamefont{A.}~\bibnamefont{Stute}}, \bibinfo {author}
  {\bibfnamefont{F.}~\bibnamefont{Dubin}}, \bibinfo {author}
  {\bibfnamefont{E.}~\bibnamefont{Phillips}}, \bibinfo {author}
  {\bibfnamefont{T.}~\bibnamefont{Monz}}, \bibinfo {author}
  {\bibfnamefont{T.}~\bibnamefont{Northup}}, \bibinfo {author}
  {\bibfnamefont{C.}~\bibnamefont{Becher}}, \bibinfo {author}
  {\bibfnamefont{T.}~\bibnamefont{Salzburger}}, \bibinfo {author}
  {\bibfnamefont{H.}~\bibnamefont{Ritsch}}, \bibinfo {author}
  {\bibfnamefont{P.}~\bibnamefont{Schmidt}},\ and\ \bibinfo {author}
  {\bibfnamefont{R.}~\bibnamefont{Blatt}},\ }%
  \bibfield{journal}{%
  \bibinfo {journal} {Applied Physics B: Lasers and Optics}\ }%
  \textbf{\bibinfo {volume} {95}},\ \bibinfo {pages} {205} (\bibinfo {year}
  {2009})%
  \bibAnnoteFile{NoStop}{russo:09}%
\bibitem{mundt:03}%
  \BibitemOpen
  \bibfield{author}{%
  \bibinfo {author} {\bibfnamefont{A.~B.}\ \bibnamefont{Mundt}}, \bibinfo
  {author} {\bibfnamefont{A.}~\bibnamefont{Kreuter}}, \bibinfo {author}
  {\bibfnamefont{C.}~\bibnamefont{Russo}}, \bibinfo {author}
  {\bibfnamefont{C.}~\bibnamefont{Becher}}, \bibinfo {author}
  {\bibfnamefont{D.}~\bibnamefont{Leibfried}}, \bibinfo {author}
  {\bibfnamefont{J.}~\bibnamefont{Eschner}}, \bibinfo {author}
  {\bibfnamefont{F.}~\bibnamefont{Schmidt-Kaler}},\ and\ \bibinfo {author}
  {\bibfnamefont{R.}~\bibnamefont{Blatt}},\ }%
  \bibfield{journal}{%
  \bibinfo {journal} {Appl. Phys. B}\ }%
  \textbf{\bibinfo {volume} {76}},\ \bibinfo {pages} {117} (\bibinfo {year}
  {2003})%
  \bibAnnoteFile{NoStop}{mundt:03}%
\bibitem{keller:04}%
  \BibitemOpen
  \bibfield{author}{%
  \bibinfo {author} {\bibfnamefont{M.}~\bibnamefont{Keller}}, \bibinfo {author}
  {\bibfnamefont{B.}~\bibnamefont{Lange}}, \bibinfo {author}
  {\bibfnamefont{K.}~\bibnamefont{Hayasaka}}, \bibinfo {author}
  {\bibfnamefont{W.}~\bibnamefont{Lange}},\ and\ \bibinfo {author}
  {\bibfnamefont{H.}~\bibnamefont{Walther}},\ }%
  \bibfield{journal}{%
  \bibinfo {journal} {New. J. Phys.}\ }%
  \textbf{\bibinfo {volume} {6}},\ \bibinfo {pages} {95} (\bibinfo {year}
  {2004})%
  \bibAnnoteFile{NoStop}{keller:04}%
\bibitem{devoe:96}%
  \BibitemOpen
  \bibfield{author}{%
  \bibinfo {author} {\bibfnamefont{R.~G.}\ \bibnamefont{DeVoe}}\ and\ \bibinfo
  {author} {\bibfnamefont{R.~G.}\ \bibnamefont{Brewer}},\ }%
  \bibfield{journal}{%
  \bibinfo {journal} {Phys. Rev. Lett.}\ }%
  \textbf{\bibinfo {volume} {76}},\ \bibinfo {pages} {2049} (\bibinfo {year}
  {1996})%
  \bibAnnoteFile{NoStop}{devoe:96}%
\bibitem{rebic:09}%
  \BibitemOpen
  \bibfield{author}{%
  \bibinfo {author} {\bibfnamefont{S.}~\bibnamefont{Rebi\'{c}}}, \bibinfo
  {author} {\bibfnamefont{J.}~\bibnamefont{Twamley}},\ and\ \bibinfo {author}
  {\bibfnamefont{G.~J.}\ \bibnamefont{Milburn}},\ }%
  \bibfield{journal}{%
  \bibinfo {journal} {Phys. Rev. Lett.}\ }%
  \textbf{\bibinfo {volume} {103}},\ \bibinfo {pages} {150503} (\bibinfo {year}
  {2009})%
  \bibAnnoteFile{NoStop}{rebic:09}%
\bibitem{mandel:qo}%
  \BibitemOpen
  \bibfield{author}{%
  \bibinfo {author} {\bibfnamefont{L.}~\bibnamefont{Mandel}}\ and\ \bibinfo
  {author} {\bibfnamefont{E.}~\bibnamefont{Wolf}},\ }%
  \emph{\bibinfo {title} {Optical Coherence and Quantum Optics}}\ (\bibinfo
  {publisher} {Cambridge University Press},\ \bibinfo {address} {London},\
  \bibinfo {year} {1995})%
  \bibAnnoteFile{NoStop}{mandel:qo}%
\end{thebibliography}
\end{document}